\documentclass[aps,notitlepage]{revtex4-1}
\usepackage{graphicx}
\usepackage{amsfonts}
\usepackage{amsmath,amssymb}
\usepackage{color}
\begin{document}

\title{Langevin dynamics, large deviations and instantons for the quasi-geostrophic
model and two-dimensional Euler equations}



\author{Freddy~Bouchet}
\email{freddy.bouchet@ens-lyon.fr}
\affiliation{Laboratoire de Physique, \'{E}cole Normale Sup\'{e}rieure de Lyon, 46 all\'{e}e d'Italie, Lyon, 69364, France}

\author{Jason~Laurie}
\email{jason.laurie@weizmann.ac.il}
\affiliation{Department of Physics of Complex Systems, Weizmann Institute of Science, 234 Herzl Street, Rehovot, 76100, Israel}

\author{Oleg~Zaboronski}
\email{o.v.zaboronski@warwick.ac.uk}
\affiliation{Mathematics Institute, Warwick University, Gibbet Hill Road, Coventry, CV4 7AL, United Kingdom}


\begin{abstract}
We investigate a class of simple models for Langevin dynamics  of turbulent
flows, including the one-layer quasi-geostrophic equation  and the two-dimensional
Euler equations. Starting from a path integral representation of the
transition probability, we compute the most probable fluctuation
paths from one attractor to any state within its basin of attraction.
We prove that such fluctuation paths are the time reversed trajectories of the relaxation
paths for a corresponding dual dynamics, which are also within the framework of quasi-geostrophic
Langevin dynamics. Cases with or without detailed balance are studied.
We discuss a specific example for which the stationary measure displays
either a second order (continuous) or a first order (discontinuous)
phase transition and a tricritical point. In situations where a  first
order phase transition is observed, the dynamics are bistable. Then, the transition
paths between two coexisting attractors are instantons (fluctuation
paths from an attractor to a saddle), which are related to the relaxation
paths of the corresponding dual dynamics. For this example, we show
how one can analytically determine the instantons and compute the transition
probabilities for rare transitions between two attractors. 
\end{abstract}

\maketitle

\section{Introduction}
\label{intro}

Many natural and experimental turbulent flows display bistable behavior,
in which one observes rare and abrupt dynamical transitions between
two attractors that correspond to very different subregions of the
phase space. The most prominent natural examples are probably the
Earth magnetic field reversals (over geological timescales), or the
Dansgaard-Oeschger events that have affected the Earth climate during
the last glacial period, and are probably due to several attractors
of the turbulent ocean dynamics \cite{Rahmstorf_2002_Nature}. Experimental
studies include examples in two-dimensional turbulence \cite{Sommeria1986,Maassen2003,Bouchet_Simonnet_2008,Gallet_Herault_2012reversals},
rotating tank experiments \cite{Weeks_Tian_etc_Swinney_Ghil_Science_1997,Schmeits_Dijkstra_bimodal_Kurosh_GulfStr_2001JPO}
related to the quasi-geostrophic dynamics of oceans (Kuroshio current
bistability \cite{Schmeits_Dijkstra_bimodal_Kurosh_GulfStr_2001JPO,Qiu_Miao_2000JPO....30.2124Q})
and atmospheres (weather regime blockings), three dimensional turbulent
flows in a Von K\'{a}rm\'{a}n geometry \cite{Ravelet_Marie_Chiffaudel_Daviaud_PRL2004},
the magnetic field reversal in MHD experiments \cite{Berhanu_etc_Fauve_2007_EPL_MagneticFieldReversal,Gallet_Herault_2012reversals},
Rayleigh-B\'{e}nard convection cells \cite{Niemela2001,Chandra2011,Sugiyama2010,Brown2006}. 

The theoretical understanding of these transitions is an extremely
difficult problem due to the large number of degrees of freedoms,
the broad spectrum of timescales and the non-equilibrium nature of
these flows. Up to now there have been an extremely limited number
of theoretical results, the analysis being mostly limited to analogies
with models with few degrees of freedom. One example with an interesting
phenomenological approach results in the clever use of symmetry
arguments in order to describe effectively the largest scales of MHD
experiments \cite{Fauve_Petrelis_2010mechanisms}. This strategy has
been fruitful in several examples in regimes close to deterministic
bifurcations, where the hypothesis of describing the turbulent flow
by few dominant modes, even if based only up to now on empirical arguments,
is likely to be relevant. In fact it has led to the prediction of
non-trivial qualitative features of the rare transitions.

The main problem is in how to develop a general theory for these phenomena?
When a complex turbulent flow switches at random times from one subregion
of the phase space to another, the first theoretical aim is to characterize
and predict the observed attractors. This is already a non-trivial
task, as no picture based on a potential landscape is available. Indeed,
this is especially tricky when the transition is not related to any
symmetry breaking. An additional theoretical challenge is in being able to compute the
transition rates between attractors. It is also often the case that
most transition paths from one attractor to another concentrate close
to a single unique path, therefore a natural objective is to compute
this most probable transition path. In order to achieve these goals,
it is convenient to think about the framework of large deviation theory,
in order to describe either the stationary distribution of the system,
or in computing the transition probabilities of the stochastic process.
In principle, we could argue that from a path integral representation
of the transition probabilities \cite{Zinn-Justin-2002-QuantumFiledTheory-Book},
and the study of its semi-classical limit in an asymptotic expansion
with a well chosen small parameter, we could derive a large deviation
rate function that would characterize the attractors and various other
properties of the system. When this semi-classical approach is relevant, one expects
a large deviation result, similar to the one obtained through the
Freidlin-Wentzell theory \cite{Freidlin_Wentzel_1984_book}. If this
notion is correct, then this would explain why these rare transitions
share many analogies with phase transitions in statistical mechanics
and stochastic dynamics with few degrees of freedom. The theoretical
issues in order to assess the validity of such a broad approach are
however numerous: what is the natural asymptotic large deviation parameter?
Why and when should the finite dimensional picture be valid? How does
one actually compute the large deviation rate function and characterize
its minima? Should one expect that the dynamics of the rare transition be well described by few degrees of freedom? And so on. Up to now, these
questions have no clear or precise answers for any meaningful turbulence problems. The aim of this paper is to make small steps in this direction.

We will study the class of models that describe two-dimensional and
quasi-geostrophic dynamics. Those are arguably the simplest class
of turbulence models for which phase transitions and bistability phenomena
exist. For simplicity, we will consider forces which are stochastic,
white in time, Gaussian noises. In previous papers, we have
given partial answers to the theoretical challenges discussed above.
For instance, for the two-dimensional stochastic Navier-Stokes equations,
we have argued \cite{Bouchet_Simonnet_2008} that in the inertial
limit (weak noise and dissipation), one should expect the invariant measure
to be concentrated close to the attractors of the inertial dynamics
(the two-dimensional Euler equations). This partially answers the issue
of characterizing the attractors, and helps us to empirically find
the bistable regimes, based on bifurcation diagrams for the inertial
dynamics. Indeed, numerical simulations showed that the Navier-Stokes
dynamics actually concentrates close to the set of attractors of the
two-dimensional Euler equations \cite{Bouchet_Simonnet_2008}, and
display bistable behavior in some parameter range. In order to develop
further the theoretical understanding, we have used stochastic averaging
techniques to describe the long time dynamics of the barotropic quasi-geostrophic
model in a regime where the main attractors are simple parallel flows
(zonal jets) \cite{Bouchet_Nardini_Tangarife_2013_Kinetic_JStatPhys}.
Moreover, this model also displays multiple attractors \cite{Bouchet_Nardini_Tangarife_2013_Kinetic_JStatPhys},
which can be studied using large deviation theory. Furthermore, we
have also developed a similar theoretical approach
for the stochastic Vlasov equations where bistability was also discussed~\cite{NardiniGuptaBouchet-2012-JSMTE,Nardini_Gupta_Ruffo_Dauxois_Bouchet_2012_kinetic}.
However these works only partially address the theoretical questions:
mainly in predicting the set of attractors and in determining the phase transitions
and bistability regimes.  However, up to now it has not been possible to explicitly compute
the transition rates and transition probabilities for these systems.

For turbulent dynamics, in inertial limits, the attractors are expected
to be subclasses of the attractors of the inertial dynamics, as we
discussed above. The natural attractors of the inertial dynamics are
those derived from the microcanonical measures, namely the macroscopic
equilibria of the Miller-Robert-Sommeria theory \cite{Robert:1990_CRAS,Miller:1990_PRL_Meca_Stat,Robert:1991_JSP_Meca_Stat,SommeriaRobert:1991_JFM_meca_Stat}
(please note the many recent contributions to the application of this
theory \cite{Venaille_2012_Bottom_Traped,Venaille_Vallis_2012JFM...709..490V,Herbert_2013_JStatMech,Herbert_Dubrulle_Chavanis_Paillard_2012statistical,Herbert_Marino_Pouquet_2013statistical,Thalabard_Dubrulle_Bouchet_2013_statistical,NasoChavanisDubrulle,NasoChavanisDubrulle2,Weichman_2012_long,Bouchet_Corvellec_JSTAT_2010,Corvellec_Bouchet_2012_condmat,BouchetPottersVaillant-2013-JSMTE}).
In essence, these microcanonical measures are characterized by an
entropy functional that is actually a large deviation rate functional
(see for instance see \cite{Michel_Robert_LargeDeviations1994CMaPh.159..195M,Boucher_Ellis_Turkington_2000_JSP}).
As explained in \cite{Bouchet:2008_Physica_D}, the related entropy
maximization is closely related to energy-Casimir variational problems.
This link highlights the possibility that energy-Casimir functionals
are natural potentials for the effective description of the largest
scales in these turbulent flows. We address this point further in the
conclusion.

The goal of this paper is to define and to study a class of Langevin
dynamics associated to energy-Casimir potentials and in the investigation
of the related stochastic process. We show that this stochastic process
is an equilibrium one, in the sense that either it verifies detailed
balance, or a generalization of the detailed balance property. In
the latter, the time reversed stochastic process is not simply
the initial process but belongs to the same class of physical model
(for instance in Langevin dynamics of particles in magnetic fields).
From this time reversal symmetry, identified at the level of the action,
we can show that the quasi-potential related to the action minimization
can be explicitly computed, and is actually the energy-Casimir functional.
Moreover, we can also explain why fluctuation trajectories (the most probable
paths to get a rare fluctuation) are time reversed relaxation trajectories
of the dual dynamics, as in classical Langevin dynamics. In situations
with bistability (when the quasi-potential has two or more local minima),
we recover the classical picture: an Arrhenius law for the transition
rate and a typical transition trajectory that follows an instanton
trajectory (the time reversed trajectory of the relaxation path of
the dual dynamics from the lowest saddle point). All these properties
are derived from the orthogonality of the Hamiltonian part of the
dynamics to the potential part, which is a consequence
of the fact that the potential is conserved under the Hamiltonian
dynamics.

We discuss a specific example where the energy-Casimir functional
leads to bistable regimes, and describe a bifurcation diagram
that includes a tricritical point (a bifurcation from a first order phase transition
to a second order phase transition). Close to the critical point,
the turbulent dynamics can be reduced to the effective dynamics involving
only a few degrees of freedom related to the null space of the potential
at the transition point, by analogy with the phenomenology of bifurcations
in deterministic systems. However, far away from the tricritical point
such a reduction does not seem to be relevant. 

These Langevin dynamics are very interesting examples of turbulent
dynamics, that fit within the classical framework of equilibrium stochastic
thermodynamics. All the recent results related to stochastic thermodynamics:
Gallavotti-Cohen fluctuation relations, relations between the entropy
production and the probability of paths, and so on, could be easily
generalized for these Langevin dynamics. Together with genuine turbulence
dynamics, they also display fascinating dynamical behavior including
phase transitions. The relevance of these dynamics for real physical
phenomena should however be questioned. As discussed in the paper
and in the conclusion, several examples of these Langevin dynamics
actually relate to physical microscopic dissipation mechanisms (linear
friction and/or viscosity), but is not true in general. When this analogy is incorrect, these dynamics should be understood, at best,
as effective models for the largest scales of the flows. All these aspects
and the resulting limitations and benefits of these model to real
flows are further discussed in the conclusion.

This Langevin dynamics approach also opens up a new set of very interesting
theoretical and mathematical issues. For instance, dynamics that involve
white in space noise, or colored noise but with vanishing related
frictions: under which conditions are the stochastic dynamics well-posed?
Would dynamics with regularized noise lead to qualitative similar
behavior? What are the necessary and sufficient conditions for the
formal computations performed in this work to be mathematically founded?
Some of these questions are related to recent advances in the mathematics
of stochastic partial differential equations \cite{kuksin2012mathematics,Kuksin_2004_JStatPhys_EulerianLimit,jaksic2012large,jaksic2013large,bessaih2012invariant,bessaih2014inviscid,gourcy2007large}.
Again, these aspects are further discussed in the conclusion.

In Section \ref{sec:Langevin-general} we discuss a general framework
for Langevin dynamics. Starting from a few hypotheses (Liouville theorem,
transversality condition, and relation between friction and noise
amplitude), we derives the time reversal symmetry properties of the
stochastic process. Section \ref{sec:The-2D-Euler-QG-dynamics} applies
this framework to two-dimensional and quasi-geostrophic turbulence
models. Section \ref{sec:Phase-transition-QG} discusses a specific
case where a tricritical point is a situation for bistability, and finally
Section \ref{sec:Conclusions} concludes by emphasizing the interest
and limitations of these Langevin models and outlining the perspectives.

\section{Langevin dynamics and equilibrium instantons}

\label{sec:Langevin-general}

The aim of this section is to describe the general framework for Langevin
dynamics. We first define Langevin dynamics in subsection \ref{sub:Langevin-dynamics},
as  stochastic, ordinary or partial, differential equations, for which
the deterministic part is composed of a vector field with a Liouville
property (conservation of phase space volume, Eq. (\ref{eq:Liouville}))
plus a potential force with potential $\mathcal{G}$. The conservative
part of the dynamics are assumed to be transverse to the gradient of
the potential (\ref{eq:Conservation_G}). The stochastic force is
defined as the derivative of a Brownian process, with a correlation function identical to that of the kernel of the potential force. 

We derive the main properties of Langevin dynamics: its invariant
measure is a Gibbs measure with potential $\mathcal{G}$. As the Langevin
dynamics is a Markov process, we can define the time reversed Markov
process, which also satisfies Langevin dynamics which is usually related to the original dynamics. We call this process the reversed, or dual Langevin dynamics.
We study this time-reversal symmetry
through the symmetry of the action, describing transition path probabilities.
Based on this symmetry, we describe the relation between relaxation
paths (most probable paths for a relaxation from any initial state
to an attractor of the system) and fluctuation paths (most probable
paths to observe a fluctuation starting from an attractor and ending
at any point of the system). As we explain, for Langevin dynamics,
fluctuation paths are time reversed trajectories of relaxation paths of the dual
dynamics.

These properties, for instance the relation between fluctuation and
relaxation paths, can be considered as a generalization of Onsager
reciprocal relations. However, they are valid for fluctuations arbitrarily
far from the main attractor, and for relaxation dynamics that do not necessarily need to be linear. Such a symmetry between the fluctuation
and relaxation paths is somehow a classical remark in statistical mechanics.
For instance, the relation between the action symmetry and detailed balance can be found in \cite{janssen1979field}, discussion of these properties can also be found in \cite{luchinsky1998analogue}, and additionally we have been told that this symmetry may be traced back to Onsager and Machlup \cite{Onsager_Machlup_1953PhRv}. Even if the basic ideas seem classical,
we do not yet know of any references where the general structure of Langevin
dynamics, and its relation with the symmetries of relaxation and fluctuation
paths are discussed. We also note an interesting discussion of this symmetry in \cite{TailleurJPA08}. This symmetry is also clearly related to the Gallavotti-Cohen fluctuation
relations \cite{Evans_Cohen_Moriss_1993PhRvL,Gallavotti_Cohen_1995_PhRvL}.

The fact that large deviation functionals can be computed explicitly
when the dynamics can be decomposed into the sum of a gradient and
a transverse part is explained in the book of Freidlin-Wentzell \cite{Freidlin_Wentzel_1984_book}.
In our problem, this transversality comes from the Hamiltonian structure
and the fact that the potential is a conserved quantity of the
Hamiltonian dynamics. As explained very clearly in \cite{BDSGJLL1},
for non-equilibrium systems, the deterministic vector field can also
be decomposed into the sum of the gradient of the quasi-potential
plus a transverse part, the transversality condition being then equivalent
to the Hamilton-Jacobi equation. Similar ideas can also be found in
works of Graham in the 1980s and 1990s (see for instance \cite{Graham1987macroscopic}).

\subsection{Langevin dynamics with potential $\mathcal{G}$ }

\label{sub:Langevin-dynamics}

We call the Langevin dynamics for the potential $\mathcal{G}$ the stochastic
dynamics given by
\begin{equation}
\frac{\partial q}{\partial t}=\mathcal{F}\left[q\right]({\bf r})-\alpha\int_{\mathcal{D}}\, C({\bf r},{\bf r}')\frac{\delta\mathcal{G}}{\delta q({\bf r}')}\left[q\right] \,{\rm d}{\bf r}'+\sqrt{2\alpha\gamma}\, \eta,
\label{eq:Langavin dynamics}
\end{equation}
where $\mathcal{F}$ satisfies a Liouville property (defined below,
Eq. (\ref{eq:Liouville})), $\mathcal{G}$ is a conserved quantity
of the dynamics defined by $\mathcal{F}$ (see Eq. (\ref{eq:Conservation_G})),
and the stochastic force $\eta$ is a Gaussian process, white in time,
with correlation function $\mathbb{E}\left[\eta({\bf r},t)\eta({\bf r}',t')\right]=C({\bf r},{\bf r}')\delta(t-t')$.
As it is a correlation function, $C$ is a symmetric
positive function, i.e. for any function $\phi$ over $\mathcal{D}$
\begin{equation}
\int_{\mathcal{D}}\int_{\mathcal{D}}\,\phi\left({\bf r}\right)\, C({\bf r},{\bf r}')\, \phi\left({\bf r}'\right)\, {\rm d}{\bf r}\, {\rm d}{\bf r}' \geq0 ,\label{eq:Correlation_positivity-1}
\end{equation}
and $C({\bf r},{\bf r}')=C({\bf r}',{\bf r})$. For simplicity, we
assume in the following that $C$ is positive definite and has an
inverse $C^{-1}$ such that 
\begin{equation*}
\int_{\mathcal{D}}\, C({\bf r},{\bf r}_{1})\, C^{-1}({\bf r}_{1},{\bf r}')\, {\rm d}{\bf r}_{1}=\delta\left({\bf r}-{\bf r}'\right).
\end{equation*}

The variable $q$ is either finite dimensional  (for instance
$q\in\mathbb{R}^{N}$), or a field (for instance a two-dimensional
field for solution of the two-dimensional Euler equations). If $q\in\mathbb{R}^{N}$,
we assume that the deterministic dynamical system 
\begin{equation}
\frac{\partial q}{\partial t}=\mathcal{F}\left[q\right],\label{eq:dynamical system}
\end{equation}
conserves the Lebesgue measure $\prod_{i=1}^{N}\, {\rm d}q_{i}$, or equivalently
that the divergence of the vector field $\mathcal{F}$ is zero:
\begin{equation}
\nabla\cdot\mathcal{F}\equiv\sum_{i=1}^{N}\frac{\partial\mathcal{F}}{\partial q_{i}}=0.\label{eq:Liouville-1}
\end{equation}
We call this property a Liouville property. If $q$ is a field (for
instance a two-dimensional vorticity or potential vorticity field,
for the two-dimensional Euler or quasi-geostrophic equations) defined over a domain
$\mathcal{D}$, we assume that a Liouville property holds, in the
sense that the formal generalization of the finite dimensional Liouville
property,
\begin{equation}
\nabla \cdot \mathcal{F}\equiv\int_{\mathcal{D}}\,\frac{\delta\mathcal{F}}{\delta q({\bf r})}\,{\rm d}{\bf r}=0,\label{eq:Liouville}
\end{equation}
is verified. We further assume that the deterministic dynamical system
(\ref{eq:dynamical system}) has $\mathcal{G}$ as a conserved quantity. Then for any $q$:
\begin{equation}
\int_{\mathcal{D}}\,\mathcal{F}\left[q\right](\mathbf{r})\frac{\delta\mathcal{G}}{\delta q({\bf r})}\left[q\right]\,{\rm d}{\bf r}=0.\label{eq:Conservation_G}
\end{equation}
This equation is a transversality property between the the vector
field $\mathcal{F}$ and the gradient of the potential $\mathcal{G}$.

These two hypotheses, Liouville (\ref{eq:Liouville}) and the conservation
of the potential (\ref{eq:Conservation_G}), are verified
if the dynamical system is Hamiltonian:
\begin{equation}
\mathcal{F}[q]=\left\{ q,\mathcal{H}\right\},
\end{equation}
with $\mathcal{G}$ being  one of its conserved quantity, for instance $\mathcal{G}=\mathcal{H}$.
We stress however that $\mathcal{G}$ does not need to be $\mathcal{H}$ in general. 

The major property of Langevin dynamics is that the stationary probability
density functional is known a-priori and is given by
\begin{equation}
P_{s}[q]=\frac{1}{Z}\exp\left(-\frac{\mathcal{G}[q]}{\gamma}\right),
\end{equation}
where $Z$ is a normalization constant. At a formal level, this can
be easily checked by writing the Fokker-Planck equation for the evolution
of the probability functional. Then the property that $P_{s}$ is stationary
readily follows from the Liouville property and the fact that
$\mathcal{G}$ is a conserved quantity for the deterministic dynamics.

\subsection{Reversed Langevin dynamics}

We consider the linear operator $I$ to be a linear involution on the space of fields $q$ ($I^{2}={\rm Id}$). Therefore, we define the reversed
Langevin dynamics, with respect to $I$, as
\begin{equation}
\frac{\partial q}{\partial t}=\mathcal{F}_{r}\left[q\right](\mathbf{r})-\alpha\int_{\mathcal{D}}\, C_{r}({\bf r},{\bf r}')\frac{\delta\mathcal{G}_{r}}{\delta q({\bf r}')}\left[q\right]\,{\rm d}{\bf r}'+\sqrt{2\alpha\gamma}\eta,\label{eq:Reverse process}
\end{equation}
where
\begin{equation}
\mathcal{F}_{r}=-I\circ\mathcal{F}\circ I,\label{eq:Fr}
\end{equation}
\begin{equation}
C_{r}=I^{+}CI,\label{eq:Cr}
\end{equation}
here $I^{+}$ is the adjoint of $I$ for the $L^{2}$ scalar product,
and 
\begin{equation}
\mathcal{G}_{r}\left[q\right]=\mathcal{G}\left[I\left[q\right]\right].\label{eq:Gr}
\end{equation}
From the properties of $\mathcal{F}$, $C$ and $\mathcal{G}$, one
can demonstrate that a Liouville property holds for $\mathcal{F}_{r}$,
that $C_{r}$ is positive definite, and that $\mathcal{G}_{r}$ is
a conserved quantity for the dynamics $\frac{\partial q}{\partial t}=\mathcal{F}_{r}\left[q\right]$ for any $q$:
\begin{equation}
\int_{\mathcal{D}}\,\mathcal{F}_{r}\left[q\right](\mathbf{r})\frac{\delta\mathcal{G}_{r}}{\delta q({\bf r})}\left[q\right]\,{\rm d}{\bf r}=0.\label{eq:Conservation_Gr}
\end{equation}
As a consequence, the reversed Langevin dynamics (\ref{eq:Reverse process})
is also Langevin. 

A very interesting case is when the deterministic dynamics is symmetric
with respect to time reversal. Then there exists a linear involution
$I$ such that
\begin{equation}
\mathcal{F}=\mathcal{F}_{r}=-I\circ \mathcal{F}\circ I.\label{eq:Time-rersal-condition-F}
\end{equation}
Moreover, if $C$ and $\mathcal{G}$ are symmetric with respect to
the involution: $C_{r}=C$ and 
\begin{equation}
\mathcal{G}_{r}=\mathcal{G},\label{eq:Time-reverseal-condition-G}
\end{equation}
then the reversed Langevin dynamics are nothing else than the original
Langevin dynamics. In this case, we say that the Langevin dynamics are
time-reversible. Simple examples of time-reversible Langevin dynamics
are the overdamped processes:
\begin{equation}
\dot{q}=-\int_{\mathcal{D}}\, C({\bf r},{\bf r}')\frac{\delta\mathcal{G}}{\delta q({\bf r}')}\left[q\right]\, {\rm d}{\bf r}'+\sqrt{2\gamma}\eta,
\end{equation}
which can be proved to be time-reversible with the involution $I={\rm Id}$, the canonical Langevin dynamics 
\begin{equation}
\begin{cases}
\dot{x}  =  p,\nonumber\\
\dot{p}  =  -\frac{{\rm d}V}{{\rm d}x}-\alpha p+\sqrt{2\alpha k_{B}T}\eta,\nonumber
\end{cases}
\end{equation}
with $I\left(x,p\right)^T=\left(x,-p\right)^T$, or the two-dimensional stochastic Euler equations:
\begin{equation}
\frac{\partial\omega}{\partial t}+{\bf v}\cdot\nabla\omega=-\alpha\int_{\mathcal{D}}\, C({\bf r},{\bf r}')\frac{\delta\mathcal{G}}{\delta\omega({\bf r}')}\, {\rm d}{\bf r}'+\sqrt{2\alpha\gamma}\eta,\quad {\rm with}\quad {\bf v}=\mathbf{e}_{z}\times\nabla\psi,\nonumber
\end{equation}
under the assumption that $\mathcal{G}$ is conserved by the Euler dynamics,
and is an even functional ($\mathcal{G}\left[-\omega\right]=\mathcal{G}\left[\omega\right]$).
For the two-dimensional Euler equations, the natural involution corresponding to
time-reversal symmetry is $I\left[\omega\right]=-\omega$. In the following,
we will also consider cases when the Langevin dynamics are not time-reversible, for instance the two-dimensional stochastic Euler equations when $\mathcal{G}$
is not even, or the quasi-geostrophic equations with topography $h(y)\neq0$.

\subsection{Path integrals, action, and time-reversal symmetry}

The Lagrangian $\mathcal{L}$ associated to the Langevin
dynamics (\ref{eq:Langavin dynamics}) is defined as 
\begin{eqnarray}
\mathcal{L}\left[q,\frac{\partial q}{\partial t}\right]&=&\frac{1}{4\alpha}\int_{\mathcal{D}}\int_{\mathcal{D}}\,\left(\frac{\partial q}{\partial t}-\mathcal{F}\left[q\right](\mathbf{r})+\alpha\int_{\mathcal{D}}C({\bf r},{\bf r}_{1})\frac{\delta\mathcal{G}}{\delta q({\bf r}_{1})}\left[q\right]\mbox{d}{\bf r}_{1}\right)\nonumber\\
&\times&C^{-1}({\bf r},{\bf r}')\left(\frac{\partial q}{\partial t}-\mathcal{F}\left[q\right](\mathbf{r}')+\alpha\int_{\mathcal{D}}\, C({\bf r}',{\bf r}_{2})\frac{\delta\mathcal{G}}{\delta q({\bf r}_{2})}\left[q\right]\mbox{d}{\bf r}_{2}\right)\,{\rm d}{\bf r}\,{\rm d}{\bf r}',\label{eq:Lagrangian}
\end{eqnarray}
and the action functional as 
\begin{equation}
\mathcal{A}_{(0,T)}\left[q\right]=\int_{0}^{T}\mathcal{L}\left[q(t),\frac{\partial q}{\partial t}(t)\right]\,{\rm d}t.\label{eq:A}
\end{equation}
Consequently, the Lagrangian of the reverse process is defined as 
\begin{eqnarray}
\mathcal{L}_{r}\left[q,\frac{\partial q}{\partial t}\right]&=&\frac{1}{4\alpha}\int_{\mathcal{D}}\int_{\mathcal{D}}\,\left(\frac{\partial q}{\partial t}-\mathcal{F}_{r}\left[q\right](\mathbf{r})+\alpha\int_{\mathcal{D}}\, C_{r}({\bf r},{\bf r}_{1})\frac{\delta\mathcal{G}_{r}}{\delta q({\bf r}_{1})}\left[q\right]\mbox{d}{\bf r}_{1}\right)\nonumber\\
&\times&C_{r}^{-1}({\bf r},{\bf r}')\left(\frac{\partial q}{\partial t}-\mathcal{F}_{r}\left[q\right](\mathbf{r}')+\alpha\int_{\mathcal{D}}\, C_{r}({\bf r}',{\bf r}_{2})\frac{\delta\mathcal{G}_{r}}{\delta q({\bf r}_{2})}\left[q\right]\mbox{d}{\bf r}_{2}\right)\,{\rm d}{\bf r}\, {\rm d}{\bf r}',
\end{eqnarray}
with the time-reversed action functional $\mathcal{A}_{r}$ defined accordingly.

Using the Onsager-Machlup formalism, we know that $P\left[q_{T},T;q_{0},0\right]$,
the transition probability to go from the state $q_{0}$ at time
$0$ to the state $q_{T}$ at time $T$, can be expressed as
\begin{equation}
P\left[q_{T},T;q_{0},0\right]=\int^{q(T)=q_T}_{q(0)=q_0}\,\mathcal{D}\left[q\right]\,{\exp}\left(-\frac{\mathcal{\mathcal{A}}}{\gamma}\right),\label{eq:Path_integral}
\end{equation}
where we have used the fact that the Jacobian
\begin{equation*}
J\left[q\right]=\left|\det\left[\frac{\delta}{\delta q({\bf r}')}\left(\dot{q}-\mathcal{F}[q]+\alpha\int_{\mathcal{D}}\, C({\bf r},{\bf r}_{1})\frac{\delta\mathcal{G}}{\delta q({\bf r}_{1})}\left[q\right]\, {\rm d}{\bf r}_{1}\right)\right]\right|,
\end{equation*}
 is formally equal to a $q$-independent constant when we interpret our stochastic partial differential equation using Ito's convention \cite{Zinn-Justin-2002-QuantumFiledTheory-Book}. This constant can be included in the definition of the functional integration measure.

For a given path $\left\{ q(t)\right\} _{0\leq t\leq T}$, we define
the reversed path by $q_{r}(t)=I\left[q(T-t)\right]$. The
main interest of the reversed process stems from the study of temporal
symmetries of the stochastic process and the remark that
\begin{equation}
\mathcal{A}\left[q_{r},T\right]=\mathcal{A}_{r}\left[q,T\right]-\left(\mathcal{G}\left[q(T)\right]-\mathcal{G}\left[q(0)\right]\right),\label{eq:Relation_A_Ar}
\end{equation}
or equivalently, using (\ref{eq:Gr}),
\begin{equation}
\mathcal{A}\left[q,T\right]=\mathcal{A}_{r}\left[q_{r},T\right]+\left(\mathcal{G}\left[q(T)\right]-\mathcal{G}\left[q(0)\right]\right).\label{eq:Relation_A_Ar_2}
\end{equation}

Let us prove this equality. Using the definition of $\mathcal{F}_{r}$,
$\mathcal{G}_{r}$ and $C_{r}$, (Eqs. (\ref{eq:Fr}-\ref{eq:Gr})), and using that 
\begin{equation*}
\frac{\delta\mathcal{G}_{r}}{\delta q(\mathbf{r})}\left[q\right]=I\frac{\delta\mathcal{G}}{\delta q(\mathbf{r})}\left[I\left[q\right]\right],\nonumber
\end{equation*}
with $I^{2}={\rm Id}$, we have 
\begin{eqnarray*}
\mathcal{L}\left[I\left[q\right],-\frac{\partial}{\partial t}I\left[q\right]\right]&=&\frac{1}{4\alpha}\int_{\mathcal{D}}\int_{\mathcal{D}}\,\left(\frac{\partial q}{\partial t}-\mathcal{F}_{r}\left[q\right](\mathbf{r})-\alpha\int_{\mathcal{D}}\, C_{r}({\bf r},{\bf r}_{1})\frac{\delta\mathcal{G}_{r}}{\delta q({\bf r}_{1})}\left[q\right]\mbox{d}{\bf r}_{1}\right)\nonumber\\
&\times&C_{r}^{-1}({\bf r},{\bf r}')\left(\frac{\partial q}{\partial t}-\mathcal{F}_{r}\left[q\right](\mathbf{r}')-\alpha\int_{\mathcal{D}}\, C_{r}({\bf r}',{\bf r}_{2})\frac{\delta\mathcal{G}_{r}}{\delta q({\bf r}_{2})}\left[q\right]{\rm d}{\bf r}_{2}\right)\,{\rm d}{\bf r}\, {\rm d}{\bf r}'.\nonumber
\end{eqnarray*}
Then, by expanding and using the conservation of $\mathcal{G}_{r}$ we
arrive to
\begin{equation*}
\mathcal{L}\left[I\left[q\right],-\frac{\partial I\left[q\right]}{\partial t}\right]=\mathcal{L}_{r}\left[q,\frac{\partial q}{\partial t}\right]-\int_{\mathcal{D}}\,\frac{\partial q}{\partial t}\frac{\delta\mathcal{G}}{\delta q({\bf r})}\,{\rm d}{\bf r},
\end{equation*}
or equivalently,
\begin{equation*}
\mathcal{L}\left[I\left[q\right],-\frac{\partial I\left[q\right]}{\partial t}\right]=\mathcal{L}_{r}\left[q,\frac{\partial q}{\partial t}\right]-\frac{{\rm d}}{{\rm d}t}\mathcal{G}\left[q\right].
\end{equation*}
Using the above formula and (\ref{eq:A}) in order to compute $\mathcal{A}\left[q_{r},T\right]$,
we obtain (\ref{eq:Relation_A_Ar}).\\

Performing the change of variable $q_{r}(t)=I\left[q(T-t)\right]$
in the path integral representation (\ref{eq:Path_integral}), and
using the action duality formula (\ref{eq:Relation_A_Ar}), we obtain
\begin{equation}
P\left[q_{T},T;q_{0},0\right]\exp\left(-\frac{\mathcal{G}\left[q_{0}\right]}{\gamma}\right)=P_{r}\left[I\left[q_{0}\right],T;I\left[q_{T}\right],0\right]\exp\left(-\frac{\mathcal{G}_{r}\left[I\left[q_{T}\right]\right]}{\gamma}\right),\label{eq:Duality-Transition-Probability}
\end{equation}
where $P_{r}$ is a transition probability for the reversed process.
We have thus obtain a relation between the transition probability
of the direct, forward, process and that of the reversed
one.

\subsection{Detailed balance for reversible processes}

If we assume that the Langevin equation is time-reversible, then the
direct and the reverse processes are the same, and the duality relation
for the transition probability implies
\begin{equation*}
P\left[q_{T},T;q_{0},0\right]\exp\left(-\frac{\mathcal{G}\left[q_{0}\right]}{\gamma}\right)=P\left[I\left[q_{0}\right],T;I\left[q_{T}\right],0\right]\exp\left(-\frac{\mathcal{G}\left[I\left[q_{T}\right]\right]}{\gamma}\right),
\end{equation*}
where it is also true that $\exp\left(-\mathcal{G}\left[I\left[q_{T}\right]\right]/\gamma\right)=\exp\left(-\mathcal{G}\left[q_{T}\right]/\gamma\right)$.
This result is the detailed balance property for the stochastic process.
When the reverse process is different from the direct process, then in general, detailed balance should not be verified.

\subsection{Steady states of the deterministic dynamics, critical points of $\mathcal{G}$,
and relaxation paths}

\subsubsection{Steady states and critical points of the potential $\mathcal{G}$}

Let us prove that any non-degenerate critical point of the potential
is also a steady state of the deterministic dynamics. This is a classical
result in mechanics, i.e. any non-degenerate critical point
of the energy is a steady state.

Extrema of the stationary PDF are critical points of the potential
$\mathcal{G}$. Such a critical point $q_{c}$ verifies
\begin{equation*}
\frac{\delta\mathcal{G}}{\delta q({\bf r})}\left[q_{c}\right]=0.
\end{equation*}
We assume that the critical point is non-degenerate, that the second
variations of $\mathcal{G}$ has no null eigenvalues. More explicitly,
the relation 
\begin{equation*}
\int_{\mathcal{D}}\,\frac{\delta^{2}\mathcal{G}}{\delta q({\bf r})\delta q({\bf r}')}\left[q_{c}\right]\,\phi(\mathbf{r}')=0,
\end{equation*}
implies that $\phi=0$. Now, we can prove that $q_{c}$ is also a
steady state of the Hamiltonian dynamics. 

We use the property that $\mathcal{G}$ is conserved. By taking the variational derivative $\delta/\delta q({\bf r})$
of Eq.(\ref{eq:Conservation_G}) we obtain that for any $q$
\begin{equation}
\int_{\mathcal{D}}\,\frac{\delta^{2}\mathcal{G}}{\delta q({\bf r}_{2})\delta q({\bf r})}\left[q\right]\mathcal{F}\left[q\right](\mathbf{r}_{2})\,{\rm d}{\bf r_{2}}+\int_{\mathcal{D}}\,\frac{\delta\mathcal{G}}{\delta q({\bf r}_{2})}\left[q\right]\frac{\delta\mathcal{F}}{\delta q({\bf r})}\left[q\right](\mathbf{r}_{2})\,{\rm d}{\bf r_{2}}=0.\label{eq:2-1}
\end{equation}
If we apply this formula at the critical point $q_{c}$, we can conclude
that
\begin{equation*}
\int_{\mathcal{D}}\,\frac{\delta^{2}\mathcal{G}}{\delta q({\bf r}_{2})\delta q({\bf r})}\left[q_{c}\right]\,\mathcal{F}\left[q_{c}\right](\mathbf{r}_{2})\, {\rm d}{\bf r_{2}}=0.
\end{equation*}
Moreover, using that $\mathcal{G}$ is non-degenerate we observe that for all $\mathbf{r}$ 
\begin{equation*}
\mathcal{F}\left[q_{c}\right](\mathbf{r})=0,
\end{equation*}
and thus $q_{c}$ is a steady state of the deterministic dynamics.

The remark that non-degenerate critical points of conserved quantity
are steady states also extends to their stability properties. Any stable
and non-degenerate minima or maxima of a conserved quantity is a
stable fixed point of the deterministic dynamics (again, think of
the energy or angular momentum in classical mechanics). These points are probably
about as old as classical mechanics. For infinite-dimensional problems, like the two-dimensional Euler equations
or other fluid mechanics problems, the issue may be more subtle. Indeed, one should be careful of possible norm inequivalence (an infinite
number of small scales can do a lot). But proofs about stability of critical
points of conserved quantities can still be obtained on a case by
case basis. For instance, we refer to the two Arnold stability theorems
for the two-dimensional Euler equations \cite{Arnold_1966}, or their generalization
to many other fluid mechanics problems \cite{Holm_Marsden_Ratiu_1998_EulerPoincare}.\\

Another important point is that from relations (\ref{eq:Fr})
and (\ref{eq:Gr}), it is clear that if $q_{s}$ is a steady state
of the deterministic dynamics, then $I\left[q_{s}\right]$ is a steady
state of the reversed dynamics, and vice-versa. Moreover, if $q_{c}$
is a critical point of the potential $\mathcal{G}$, then $I\left[q_{c}\right]$
will be a critical point of $\mathcal{G}_{r}$. The stability properties
(minima, global minima, local minima, number of unstable directions,
and so on) of $q_{c}$, with respect to the minimization of $\mathcal{G}$,
will agree with the stability properties of $I\left[q_{c}\right]$ with respect to the minimization of $\mathcal{G}_{r}$.

\subsubsection{Relaxation dynamics and Lyapunov functionals}

We define a relaxation path to be a solution of the relaxation dynamics:
\begin{equation}
\frac{\partial q}{\partial t}=\mathcal{F}\left[q\right](\mathbf{r})-\alpha\int_{\mathcal{D}}\, C({\bf r},{\bf r}')\frac{\delta\mathcal{G}}{\delta q({\bf r}')}\left[q\right]\,{\rm d}{\bf r}'.\label{eq:Relaxation_path}
\end{equation}

For any relaxation path $q(t)$, using the property that $\mathcal{G}$ is conserved
by the inertial dynamics we can easily prove that 
\begin{equation*}
\frac{\rm d}{{\rm d}t}\mathcal{G}\left[q(t)\right]=-\alpha\int_{\mathcal{D}}\, C({\bf r},{\bf r}')\frac{\delta\mathcal{G}}{\delta q({\bf r}')}\frac{\delta\mathcal{G}}{\delta q({\bf r})}\,{\rm d}{\bf r}\,{\rm d}{\bf r}'\leq0,
\end{equation*}
where we have used the positive definiteness of $C$ for establishing the inequality.
Thus, we  can conclude that $\mathcal{G}$ is a Lyapunov functional for
the relaxation dynamics. 

From this, we state that any minima of the potential is stable for the relaxation dynamics.

\subsection{Action minima, relaxation paths of the dual dynamics and instantons}
\label{sub:Instantons-Relaxation-Paths}

We consider action minima, subjected to fixed boundary conditions
\begin{equation}
A_{(0,T)}\left[q_{0},q_{T}\right]=\min_{\left\{ q\,\left|\,q(0)=q_{0},\  q(T)=q_{T}\right\}\right.}\mathcal{A}_{(0,T)}\left[q\right].\label{eq:Action-minima}
\end{equation}
This variational problem important for many questions. For instance,
it describes the most probable path to go from state $q_{0}$ to state $q_{T}$.
Moreover, as will be discussed in the next section, it will be useful
in order to describe large deviation results.

From the definition of the action (\ref{eq:Lagrangian}-\ref{eq:A}),
and as $C$ is positive definite, it is clear that 

\begin{equation*}
A_{(0,T)}\left[q_{0},q_{T}\right]\geq0.
\end{equation*}
Furthermore, using the action duality relation given by Eq. (\ref{eq:Relation_A_Ar_2}), we
also conclude that 
\begin{equation}
A_{(0,T)}\left[q_{0},q_{T}\right]\geq\mathcal{G}\left[q_{T}\right]-\mathcal{G}\left[q_{0}\right].\label{eq:Inequality_A_DeltaG}
\end{equation}

It is self-evident from the definition of the relaxation paths (\ref{eq:Relaxation_path}),
and from the structure of the action functional (\ref{eq:Lagrangian}-\ref{eq:A})
that a relaxation path has zero action. This should be physically intuiative,
as no noise is needed for the system to follow such a path. Then, if
there exists a relaxation path between $q_{0}$ and $q_{T}$ taking
time $T$, ($\left\{ q(t)\right\} _{0\leq t\leq T}$ such that $q(0)=q_{0}$
and $q(T)=q_{T}$), we deduce that 
\begin{equation*}
A_{(0,T)}\left[q_{0},q_{T}\right]=0.
\end{equation*}

Similarly, using the duality relation (\ref{eq:Relation_A_Ar_2}),
if there exists a relaxation path for the reversed dynamics between
$I\left[q_{T}\right]$ and $I\left[q_{0}\right]$, we surmise that 
\begin{equation*}
A_{(0,T)}\left[q_{0},q_{T}\right]=\mathcal{G}\left[q_{T}\right]-\mathcal{G}\left[q_{0}\right].
\end{equation*}
This is an important statement. Indeed, the reversed dynamics has properties
very similar to that of the original dynamics (it has the same fixed points,
the same attractors, and the same saddles up to the application of the
involution $I$), but in the argument above, we see that the final
and end-points of the relaxation paths have been exchanged from $q_{0}$
and $q_{T}$ to $I\left[q_{T}\right]$ and $I\left[q_{0}\right]$ respectively.
This will be especially useful when the starting point is one of the
local minima of the potential $\mathcal{G}$, and thus one of the
attractors of the reversed dynamics.

Consider now the case when $q_{0}$ is a local minima of $\mathcal{G}$.
Then as it is also an attractor of the relaxation dynamics, no non-trivial
relaxation path will start at $q_{0}$. But, for all $q_{T}$ inside
the basin of attraction of $q_{0}$, there
exists a relaxation path from $q_{T}$ to $q_{0}$. Generically, this
path will take an infinite amount of time $T=\infty$, e.g.  if there
is an exponential relaxation. Consequently, there will also be a relaxation path
for the dual dynamics from $I\left[q_{T}\right]$ to $I\left[q_{0}\right]$
taking infinite time. 

Therefore, for relaxation dynamics, we have that for all $q_{T}$ in the basin of attraction
of an local minima of $q_{0}$
\begin{equation*}
A_{(-\infty,0)}\left[q_{0},q_{T}\right]=\mathcal{G}\left[q_{T}\right]-\mathcal{G}\left[q_{0}\right].
\end{equation*}
For many problems, e.g. when one considers the stationary distribution, the action minima
$A_{(-\infty,0)}\left[q_{0},q_{T}\right]$ becomes an essential quantity. 

If $q_{T}$ is in basin of attraction of $q_{1}\neq q_{0}$, then
as there exists a relaxation path from $q_{1}$ to $q_{T}$,
we can infer that
\begin{equation*}
A_{(-\infty,0)}\left[q_{0},q_{T}\right]=A_{(-\infty,0)}\left[q_{0},q_{1}\right].
\end{equation*}
Moreover, it is easily understood that the action minima will correspond
to the relaxation trajectory, in the reversed dynamics, from the saddle
$q_{s}(q_{0},q_{1})$ that belongs to the closure of the basin of
attractions of both $q_{0}$ and $q_{1}$, with the smallest possible
value of the potential $\mathcal{G}\left[q_{s}(q_{0},q_{1})\right]$.
Hence, if $q_{T}$ is within the basin of attraction of $q_{1}$ we have
\begin{equation*}
A_{(-\infty,0)}\left[q_{0},q_{T}\right]=A_{(-\infty,0)}\left[q_{0},q_{1}\right]=A_{(-\infty,\infty)}\left[q_{0},q_{s}(q_{_{0},}q_{1})\right]=\mathcal{G}\left[q_{s}(q_{0},q_{1})\right]-\mathcal{G}\left[q_{0}\right].
\end{equation*}

Ultimately, the minimizers of the action, between local minima
of the potential and saddles, of infinite time, are immensely important.
These trajectories are called instantons. As it should be obvious from the
previous discussion, instantons for Langevin dynamics are the
reversed time trajectories of relaxation paths of the reversed dynamics. Instantons
are thus fluctuation paths for the Langevin dynamics. More explicitly,
if $\left\{ q_{r}(t)\right\} _{-\infty\leq t\leq\infty}$ is a relaxation
path for the reversed dynamics between a saddle $I\left[q_{s}\right]$
and the attractor $I\left[q_{0}\right]$, then the instanton between
$q_{0}$ and $q_{s}$ is given by $\left\{ I\left[q_{r}(-t)\right]\right\} _{-\infty\leq t\leq\infty}$.
As instantons are the most probable fluctuation paths between attractors and saddles, they require an infinite amount of time to leave the attractor and an infinite amount of 
time to converge to the saddle.  Moreover, they are degenerate, in the sense that
if $\left\{ q_{r}(t)\right\} _{-\infty\leq t\leq\infty}$ is an instanton,
then for any $\tau$, $\left\{ q_{r}(t+\tau)\right\} _{-\infty\leq t\leq\infty}$
is also an instanton.

\subsection{Large deviations, Freidlin-Wentzell theory and entropic effects}

Up to now, we have discussed only the symmetry properties of the action functional
(\ref{eq:A}) and of the action minima (\ref{eq:Action-minima}).
In the limit of small noise, $\gamma\rightarrow0$ (see subsection (\ref{sub:Langevin-dynamics})),
one directly observes, from the path integral representation of the transition
probability (\ref{eq:Path_integral}) that the minima of the action
will play a crucial role. Indeed, the path integral will then be seen
as a Laplace integral, and a Laplace principle will be used in order
to derive a large deviation result
\begin{equation}
P\left[q_{T},T;q_{0},0\right]\underset{\gamma\rightarrow 0}{=}\exp\left(-\frac{A_{(0,T)}\left[q_{0},q_{T}\right]}{\gamma}+{\rm o}\left(\frac{1}{\gamma}\right)\right),\label{eq:Large_Deviations}
\end{equation}
where $A_{(0,T)}\left[q_{0},q_{T}\right]=\min_{\left\{ q\,\left|\,q(0)=q_{0},\ q(T)=q_{T}\right.\right\}}\mathcal{A}_{(0,T)}\left[q\right]$,
and where ${\rm o}\left(1/\gamma\right)$ are subdominant
contributions. Physicist, through explicit computations, have discussed
many examples where this Laplace principle may or may not be correct for
small $\gamma$. In quantum mechanics, evaluations of path integrals
in the limit of small $\hbar$, or in the WKB approximation, which also involves the evaluation of path integrals through a saddle point approximation.
On the mathematical side, the study of sufficient hypotheses in order
to rigorously prove such large deviation results (\ref{eq:Large_Deviations})
is one of the main aspects of Freidlin-Wentzell theory \cite{Freidlin_Wentzel_1984_book}.
Roughly speaking, Freidlin and Wentzell proved that for finite
dimensional stochastic dynamics, under generic hypotheses, a large
deviation result actually holds.

However, we draw the attention of the reader to the fact that for
infinite dimensional field equations, e.g turbulence models,
a large deviation result (\ref{eq:Large_Deviations}) is far from
obvious in the limit of small $\gamma$. It may be expected to be
true if, for instance, if the degrees of freedom at the smallest scales can be proven
to have a negligible effect  upon the dynamics, such that it is qualitatively
similar to an effective finite dimensional system. For the turbulence
model we present here, such a property is not obvious
at all. Studying this issue in general is an extremely difficult task.  The path integral taken over
Gaussian fluctuations around the critical point is given by the determinant of the second variation
of the action functional and this determinant is typically infinite for infinitely many degrees of freedom. Therefore it requires a regularization which can either lead to a renormalization of constants in (\ref{eq:Large_Deviations}) or to a completely different answer.  This problem goes beyond the scope of this paper, however, we will return to this discussion for a specific case in the conclusion.

\section{The two-dimensional Euler and quasi-geostrophic equilibrium dynamics\label{sec:The-2D-Euler-QG-dynamics}}

In this section, we apply the formalism outlined previously to
turbulence models. We explain why the two main hypotheses of Langevin
dynamics (Liouville property and conservation of the potential related
to the transversality condition) are verified. We assume that the
kernel in front of the gradient part and the noise autocorrelation
are identical.
Then all of the time-reversal properties and  the Lyapunov properties discussed
in the previous section apply to these turbulence models.

An interesting aspect, explained below, is that depending on the properties
of the potential $\mathcal{G}$ (even or not), and of the model (with
or without topography), the Langevin dynamics can be either symmetric
under time reversal or not.

We consider the Langevin dynamics associated to the quasigeostrophic
equations in a periodic domain $\mathcal{D}=[0,2\pi l_{x})\times[0,2\pi)$
with aspect ratio $l_{x}$ to be given as
\begin{eqnarray}
\frac{\partial q}{\partial t}+\mathbf{v}\left[q-h\right]\cdot\mathbf{\nabla}q & = & -\alpha\int_{\mathcal{D}}\, C({\bf r},{\bf r}')\frac{\delta\mathcal{G}}{\delta q({\bf r}')}\,{\rm d}{\bf r}'+\sqrt{2\alpha\gamma}\eta,\label{eq:beta-plane}\\
\mathbf{v}=\mathbf{e}_{z}\times\mathbf{\nabla}\psi,\quad \omega&=&\Delta\psi, \quad q=\omega+h(\mathbf{r}),
\end{eqnarray}
with potential $\mathcal{G}.$ The stochastic force $\eta$ is a Gaussian
process, white in time, with correlation function $\mathbb{E}\left[\eta(\mathbf{r},t)\eta(\mathbf{r}',t')\right]=C(\mathbf{r},\mathbf{r}')\delta(t-t')$.
The potential $\mathcal{G}$ and the assumption of Langevin dynamics
are discussed in section \ref{sub:Conserved-quantity-and-Liouville}.
Moreover, the topography $h(\mathbf{r})$ is such that $\int_{\mathcal{D}}\, h\left(\mathbf{r}\right)\, {\rm d}{\bf r}=0$.
We consider $G$ to be the Green's function of the Laplacian operator ($G=\Delta^{-1}$)
for doubly periodic functions with zero averages. Then, the equations relating  the potential vorticity $q$, the stream function $\psi$,
and the velocity are inverted as
\begin{equation*}
\psi(\mathbf{r})=\int_{\mathcal{D}}\, G\left(\mathbf{r},{\bf r}'\right)\left[q({\bf r}')-h({\bf r}')\right]\, {\rm d}{\bf r}',
\end{equation*}
and 
\begin{equation}
\mathbf{v}\left[\omega\right](\mathbf{r})=\int_{\mathcal{D}}\,\mathbf{e}_{z}\times\nabla_{\mathbf{r}'}G\left(\mathbf{r},{\bf r}'\right)\omega(\mathbf{r}')\, {\rm d}{\bf r}',\label{eq:inversion-velocity}
\end{equation}
respectively. Here, $\mathbf{v}\left[\omega\right]$ is the operator that
allows us to compute the velocity from the vorticity. When $h=0$, these dynamics correspond to the two-dimensional Euler equilibrium dynamics.

\subsection{Conserved quantity and Liouville property\label{sub:Conserved-quantity-and-Liouville}}

From the velocity-vorticity relationship, it is easily checked that the kinetic energy can be expressed as 
\begin{equation}
\mathcal{E}=-\frac{1}{2}\int_{\mathcal{D}}\,\left[q-h\left(\mathbf{r}\right)\right]\psi\, {\rm d}\mathbf{r}=\frac{1}{2}\int_{\mathcal{D}}\,\left(\nabla\psi\right)^{2}\, {\rm d}\mathbf{r},\label{eq:Energy-QG}
\end{equation}
and, for any sufficiently smooth real function $s$, the Casimir functionals are defined as
\begin{equation*}
\mathcal{C}_{s}=\int_{\mathcal{D}}\, s(q)\, {\rm d}\mathbf{r},
\end{equation*}
which are all conserved quantities of the deterministic quasi-geostrophic dynamics (Eqs. (\ref{eq:beta-plane}) for $\alpha=0$). For any $s$,
and any $\beta$ the functional 
\begin{equation*}
\mathcal{G}=\mathcal{C}_{s}+\beta\mathcal{E},
\end{equation*}
 will be the correct potential for Langevin dynamics.

Moreover, as the deterministic equations (Eqs. (\ref{eq:beta-plane})
for $\alpha$ =0) essentially correspond to a transport equation by a divergence-less
velocity field, the Liouville property (\ref{eq:Liouville}) is formally
verified
\begin{equation*}
\nabla\cdot\mathcal{F}\equiv-\int_{\mathcal{D}}\,\mathbf{v}\left[q-h\right]\cdot\mathbf{\nabla}q\,{\rm d}\mathbf{r}=-\int_{\mathcal{D}}\,\nabla\cdot\left(\mathbf{v}\left[q-h\right]q\right)\,{\rm d}\mathbf{r}=0.
\end{equation*}

Then the formalism of section \ref{sec:Langevin-general} applies
with $\mathcal{F}\left[q\right]=-\mathbf{v}\left[q-h\right]\cdot\nabla q$.

\subsection{Reversed dynamics and detailed balance}

For the two-dimensional Euler or quasi-geostrophic equations, the relevant involution
corresponding to a time reversal is 
\begin{equation*}
I\left[q\right]=-q.
\end{equation*}
Using (\ref{eq:Fr}-\ref{eq:Gr}) we conclude that
\begin{equation*}
\mathcal{F}_{r}\left[q\right]=\mathbf{v}\left[q+h\right]\cdot\nabla q,
\end{equation*}
$C_{r}=C$ and
\begin{equation*}
\mathcal{G}_{r}\left[q\right]=\mathcal{G}\left[-q\right].
\end{equation*}

From these equations, we observe that for the two-dimensional Euler equations ($h=0$),
$\mathcal{F}_{r}=\mathcal{F}$, and thus we conclude that the dynamics
are time-reversible (see Eq. (\ref{eq:Fr})). The time reversibility
condition on $\mathcal{G}$ (see Eq. (\ref{eq:Time-reverseal-condition-G}))
imposes that the potential $\mathcal{G}$ must be even. There we have two cases:
\begin{enumerate}
\item For the two-dimensional Euler equations with an even potential $\mathcal{G}$, the Langevin
dynamics are time-reversible and detailed balance is verified.
\item When either $h\neq0$ (quasi-geostrophic) or when $\mathcal{G}$ is not
even, then the Langevin dynamics are not time-reversible. The original dynamics are conjugated
to another Langevin dynamics where $h$ has to be replaced by $-h$
and $\mathcal{G}$ by $\mathcal{G}_{r}\left[q\right]=\mathcal{G}\left[-q\right]$.
In this case, detailed balance is not verified.
\end{enumerate}

\subsection{Instanton equation\label{sub:Instanton-equation-QG}}

As discussed in section \ref{sec:Langevin-general}, the instantons
from one attractor to a saddle are given by the reverse of the relaxation
paths of the corresponding reversed dynamics. From (\ref{eq:Relaxation_path})
applied to the case where $\mathcal{F}_{r}\left[q\right]=\mathbf{v}\left[q+h\right]\cdot\nabla q$,
and $\mathcal{G}_{r}\left[q\right]=\mathcal{G}\left[-q\right]$, we
determine that the equation of these relaxation paths is 
\begin{equation}
\frac{\partial q}{\partial t}+\mathbf{v}\left[q+h\right]\cdot\nabla q=-\alpha\int_{\mathcal{D}}\, C({\bf r},{\bf r}')\frac{\delta\mathcal{G}}{\delta q({\bf r}')}\left[-q\right] {\rm d}{\bf r}'.\label{eq:relaxation-paths-QG}
\end{equation}

\subsection{Energy, enstrophy, and energy-enstrophy ensembles and physical dissipation\label{sub:Energy-enstrophy-physicaldiscussion}}

In this subsection, we consider the special case when the potential is given in the following form 
\begin{equation}
\mathcal{G}=\int_{\mathcal{D}}\,\frac{q^{2}}{2}\,{\rm d}\mathbf{r}+\beta\mathcal{E}.\label{eq:Energy-Enstrophy}
\end{equation}
This structure is referred to as the potential enstrophy ensemble (when $\beta=0$), the
enstrophy ensemble (when $\beta=0$ and $h=0$), or generally as the energy-enstrophy
ensemble. The properties of the corresponding invariant measures have been discussed on a number of occasions, starting with the works of Kraichnan \cite{Kraichnan_Motgommery_1980_Reports_Progress_Physics}
in the case of Galerkin truncations of the dynamics, and for some cases without
discretization, see for instance \cite{Bouchet_Corvellec_JSTAT_2010}
and references therein. 

For specific choices of the potential $\mathcal{G}$ and of the kernel
$C$, the friction term can also be identified with a classical physical
dissipation mechanism. For instance, if $C({\bf r},{\bf r}')=\Delta\delta(\mathbf{r}-\mathbf{r}')$,
and the potential takes the form of (\ref{eq:Energy-Enstrophy}), then
the dissipative term on the right hand side of (\ref{eq:beta-plane})
is 
\begin{equation*}
-\alpha\int_{\mathcal{D}}\, C({\bf r},{\bf r}')\frac{\delta\mathcal{G}}{\delta q({\bf r}')}\left[q\right]\,{\rm d}{\bf r}'=\alpha\Delta q-\alpha\beta q,
\end{equation*}
which leads to a diffusion type dissipation with viscosity $\alpha$ and
a linear friction with friction parameter $\alpha\beta$. Such a linear
friction can model the effects of three-dimensional
boundary layers on the quasi two-dimensional bulk vorticity, that appear in
experiments with a very large aspect ratio, rotating tank experiments,
or soap film experiments.

The fact that for the enstrophy ensemble, the quasi-potential is simply the
enstrophy, the relaxation and fluctuation paths can be easily
computed explicitly in many scenarios, as is discussed in \cite{Bouchet_Laurie_2012_Onsager_Japan}.

For the majority of the other cases, the dissipative term on the right hand
side of (\ref{eq:beta-plane}) cannot be identified as a microscopic
dissipation mechanism nor as a physical mechanism.  There is however another
possible interpretation of this kind of friction term. As explained
in \cite{Bouchet:2008_Physica_D}, entropy maxima subjected to constraints
related to the conservation of energy and the distribution of vorticity,
are also extrema of energy-Casimir functionals.
By analogy with the Allen-Cahn equation in statistical mechanics,
that uses the free energy as a potential, it seems reasonable to describe
the largest scales of turbulent flows as evolving through a gradient
term of the energy-Casimir functional. Such models have been considered
in the past (see, for example \cite{Chavanis_Generalized_Entropy_2003,ChavanisEPJB2009}
and references therein). At this stage, this should be considered
as a phenomenological approach, as no
clear theoretical results exist to support this view. 

\section{Phase transition and instantons between zonal flows in the barotropic
quasi-geostrophic equations\label{sec:Phase-transition-QG}}

In order to fully determine the quasi-geostrophic Langevin dynamics
(\ref{eq:beta-plane}), we need to specify the topography function
and the potential $\mathcal{G}$. Given the infinite number
of conserved quantities for the quasi-geostrophic dynamics, there
are many possible choices. We are interested in the description of the phenomenology
of phase transitions and instanton theory in situations of first
order transitions.  Therefore, we will illustrate such a phenomenology through
two examples. 

For the first example, we choose a topography give by $h\left(\mathbf{r}\right)=H\cos\left(2y\right)$,
such that 
\begin{equation*}
q=\Delta\psi+H\cos\left(2y\right),
\end{equation*}
and consider the potential 
\begin{equation}
\mathcal{G}=\mathcal{C}+\beta\mathcal{E},\label{eq:Potential-Zonal-Transitions-1}
\end{equation}
with energy (\ref{eq:Energy-QG}), $\beta$ the inverse temperature,
and where $\mathcal{C}$ is the Casimir functional 
\begin{equation}
\mathcal{C}=\int_{\mathcal{D}}\,\frac{q^{2}}{2}-a_{4}\frac{q^{4}}{4}+a_{6}\frac{q^{6}}{6}\, {\rm d}\mathbf{r},\label{eq:C-a4}
\end{equation}
where we assume that $a_{6}>0$.

\subsection{Zonal phase transitions\label{sub:Zonal-phase-transitions}}

We first consider the structure of the minima of the potential $\mathcal{G}$
(\ref{eq:Potential-Zonal-Transitions-1}), and then their bifurcations
when the parameters $\epsilon$ and $a_{4}$ are changed, where $\epsilon$
is defined by 
\begin{equation*}
\beta=-1+\epsilon.
\end{equation*}
At low positive temperature ($\beta\rightarrow\infty$), we expect
to observe energy minima, which correspond to $\psi=0$ and $q=H\cos\left(2y\right)$.
As the energy is convex, for positive $\beta$ and small enough $a_{4}$,
both $\mathcal{C}$ and $\beta\mathcal{E}$ will also be convex. 
Henceforth, we expect that $\mathcal{G}$ will contain an unique global minimum and
no local minima. For large enough $\beta$, this equilibrium state
will be dominated by the topographic effect. For small negative $\beta$,
the change of convexity of $\beta\mathcal{E}$ from convex to concave
will not change this picture. However, for smaller $\beta$ (more negative
and higher absolute value), we expect a phase transition to occur
as the potential $\mathcal{G}$ will become locally concave. If $a_{4}>0$, with
sufficiently large values, this will be a first order phase transition.
If $a_{4}<0$ with sufficiently large values, this will be a second
order phase transition.

When $H=0$, a bifurcation occurs for $\beta=-1$ ($\epsilon=0$) and
$a_{4}=0$, as can be easily checked (see {\cite{Corvellec_Bouchet_2012_condmat}).
This bifurcation is due to the vanishing of the Hessian
at $\beta=-1$ ($\epsilon=0$) and $a_{4}=0$. As discussed in many
papers \cite{ChavanisSommeria:1996_JFM_Classification,Venaille_Bouchet_PRL_2009,BouchetVenaille-PhysicsReport,Corvellec_Bouchet_2012_condmat},
for the quadratic Casimir functional $\mathcal{C}_{2}=\int_{\mathcal{D}}\,q^{2}/2\, {\rm d}{\bf r}$,
the first bifurcation involves the eigenfunction of $-\Delta$ with
the lowest eigenvalue. If we assume that the aspect ratio $l_{x}$
(defined just before equation (\ref{eq:beta-plane})) satisfies $l_{x}<1$,
then the smallest eigenvalue is the one corresponding to the zonal
mode proportional to $\cos\left(y\right)$. Because we are interested by transitions
between two zonal states, we assume from now on that $l_{x}<1$. 

For non-zero, but sufficiently small, $H$ there will still be a bifurcation
for $\epsilon$ and $a_{4}$ close to zero. This is the regime that we wish to consider. The null space of the Hessian is spanned by eigenfunctions
$\cos\left(y\right)$ and $\sin\left(y\right)$, therefore as a consequence, for small enough $\epsilon$,
$a_{4}$ and $H$, we expect that the bifurcation can be described by a normal
form involving only the projection of the field $q$ onto the null
space. Hence, we decompose the fields into a contribution arising through its projection onto this
null space and its orthogonal complement:  
\begin{equation}
\psi=A\cos\left(y\right)+B\sin\left(y\right)+\psi'\label{eq:psi-A}
\end{equation}
where $\int_{\mathcal{D}}\exp\left(iy\right)\, \psi'(\mathbf{r})\, {\rm d}{\bf r} =0.$
Then 
\begin{equation}
q=-A\cos\left(y\right)-B\sin\left(y\right)+q',\label{eq:psi-B}
\end{equation}
with $\int_{\mathcal{D}}\,\exp\left(iy\right)\, q'(\mathbf{r}) \,{\rm d}{\bf r}=0$.
The fact that the bifurcation can be described by a normal form over
the null space of the Hessian can be expected on a general basis.
It can actually be justified by using Lyapunov-Schmidt reduction,
as performed and explained in \cite{Corvellec_Bouchet_2012_condmat}
for a number of examples for the two-dimensional Euler and quasi-geostrophic equations.
Then all other degrees of freedoms describing the minima $q_{c}$
of $\mathcal{G}$ are slaved to $A$ and $B$, in the sense that they
can be simply expressed as functions of $A$ and $B$ themselves. Even though the
following example is not treated in the paper \cite{Corvellec_Bouchet_2012_condmat},
it would not be difficult.  Therefore, we omit the details of the Lyapunov-Schmidt reduction here for simplicity.
Instead, we rather propose a more heuristic discussion.

Our strategy, will be in treating the problem perturbatively by assuming that $\epsilon\ll1$,
$\epsilon a_{6}\ll a_{4}^{2}$, and $a_{4}H^{2}\ll\epsilon$ (note
that it implies that $a_{6}H^{4}\ll\epsilon$). We make these assumptions
in order to get an explicit description of the phase transition.
However, it is important to understand that the theory that predicts
the transition rates and the instantons does not depend on these
assumptions, and that the same phenomenology will remain valid beyond
the perturbative regime. We will assume that $\psi'$ and $q'$
are first order corrections in all of the three perturbation parameters.
By rewriting the potential $\mathcal{G}$, taking into account only the leading
order contributions, and using Eqs. (\ref{eq:Energy-QG}), (\ref{eq:C-a4})
and (\ref{eq:psi-A}-\ref{eq:psi-B}), we get after some straightforward computations that
\begin{equation*}
\mathcal{E}=\pi^{2}l_{x}\left(A^{2}+B^{2}\right)+\frac{1}{2}\int_{\mathcal{D}}\,\left[H\cos(2y)-q'\right]\psi' \, {\rm d}\mathbf{r},
\end{equation*}
and 
\begin{equation*}
\mathcal{G}=\pi^{2}l_{x}\mathcal{G}_{0}(A,B)+\mathcal{G}_{1}(A,B)\left[q'\right]+{\rm lower~order~terms},
\end{equation*}
with 
\begin{equation*}
\mathcal{G}_{0}(A,B)=\epsilon\left(A^{2}+B^{2}\right)-\frac{3a_{4}}{8}\left(A^{2}+B^{2}\right)^{2}+\frac{5a_{6}}{24}\left(A^{2}+B^{2}\right)^{3}+\mathcal{O}\left(\epsilon a_{4}\right),
\end{equation*}
and
\begin{eqnarray}
\mathcal{G}_{1}(A,B)\left[q'\right]&=&\frac{\epsilon-1}{2}\int_{\mathcal{D}}\,\left[H\cos\left(2y\right)-q'\right]\psi'\,{\rm d}\mathbf{r}\nonumber\\
&&+\frac{1}{2}\int_{\mathcal{D}}\, q'^2\left\{1-3a_{4}\left[A\cos(y)+B\sin(y)\right]^{2}+5a_{6}\left[A\cos(y)+B\sin(y)\right]^{4}\right\} \,{\rm d}\mathbf{r}.\label{eq:G1}
\end{eqnarray}
We further assume that $a_{4}A^{2}\ll\epsilon$, $a_{6}A^{4}\ll\epsilon$
and $\epsilon\ll1$. Then. the leading order terms are obtained from
the minimization of the first integral and 
\begin{equation*}
\psi'=\left[\frac{H}{3}\cos(2y)\right]\left[1+\mathcal{O}\left(\epsilon\right)+\mathcal{O}\left(a_{4}A^{2}\right)+\mathcal{O}\left(a_{6}A^{4}\right)\right],
\end{equation*}
or equivalently 
\begin{equation*}
q'=-\frac{H}{3}\cos(2y)\left[1+\mathcal{O}\left(\epsilon\right)+\mathcal{O}\left(a_{4}A^{2}\right)+\mathcal{O}\left(a_{6}A^{4}\right)\right].
\end{equation*}
We use this expression in order to compute the leading order contributions
to $G_{1}(A,B)=\min_{q'}\mathcal{\, G}_{1}(A,B)\left[q'\right]$.
After lengthy but straightforward computations, we get the leading
order contribution to be
\begin{equation*}
G_{1}=\min_{q'}\mathcal{G}_{1}=-\frac{{H^{2}}}{3}-\frac{\pi^{2}l_{x}a_{4}H^{2}}{6}\left(A^{2}+B^{2}\right)+\frac{5\pi^{2}l_{x}a_{6}H^{2}}{144}\left[5\left(A^{2}+B^{2}\right)^{2}+2\left(A^{2}-B^{2}\right)^{2}\right],
\end{equation*}
and subsequently we obtain
\begin{equation}
\min_{q}\mathcal{G}=\min_{(A,B)}\,\pi^{2}l_{x}G(A,B)\label{eq:Relation-G-GAB}
\end{equation}
with $G$ given at leading order by 
\begin{eqnarray}
G(A,B)&=&-\frac{{H^{2}}}{3}+\left(\epsilon-\frac{a_{4}H^{2}}{6}+\frac{5a_{6}H^{4}}{216}\right)\left(A^{2}+B^{2}\right)\nonumber\\
&&+\left(-\frac{3a_{4}}{8}+\frac{25a_{6}H^{2}}{144}\right)\left(A^{2}+B^{2}\right)^{2}+\frac{5a_{6}}{24}\left(A^{2}+B^{2}\right)^{3}+\frac{5a_{6}H^{2}}{72}\left(A^{2}-B^{2}\right)^{2}.\label{eq:critical_points}
\end{eqnarray}
$G(A,B)$ is the normal form that describes the phase transition in the limit
$a_{4}A^{2}\ll1$, and $a_{6}A^{4}\ll1$ and $\epsilon\ll1$.

The fact that $G$ is a normal form for small enough $a_{4}$, $a_{6}$,
and $H$, implies that the gradient of $\mathcal{G}$ in the directions
transverse to $q=A\cos\left(y\right)+B\sin\left(y\right)$ are much steeper than the gradient
of $G$. A more complete derivation could easily be performed along the lines discussed in \cite{Corvellec_Bouchet_2012_condmat}.

\begin{figure}
\begin{center}
\includegraphics[width=0.37\columnwidth]{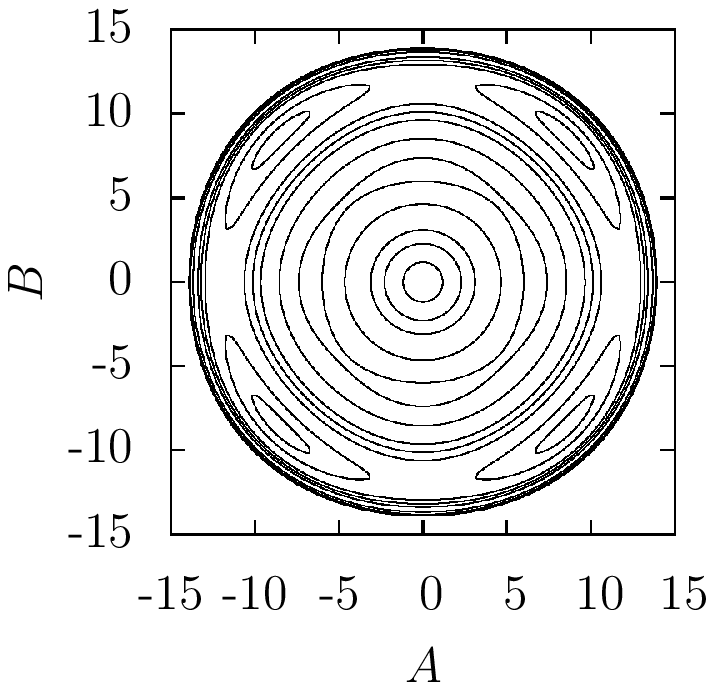}
\hfill
\includegraphics[width=0.58\columnwidth]{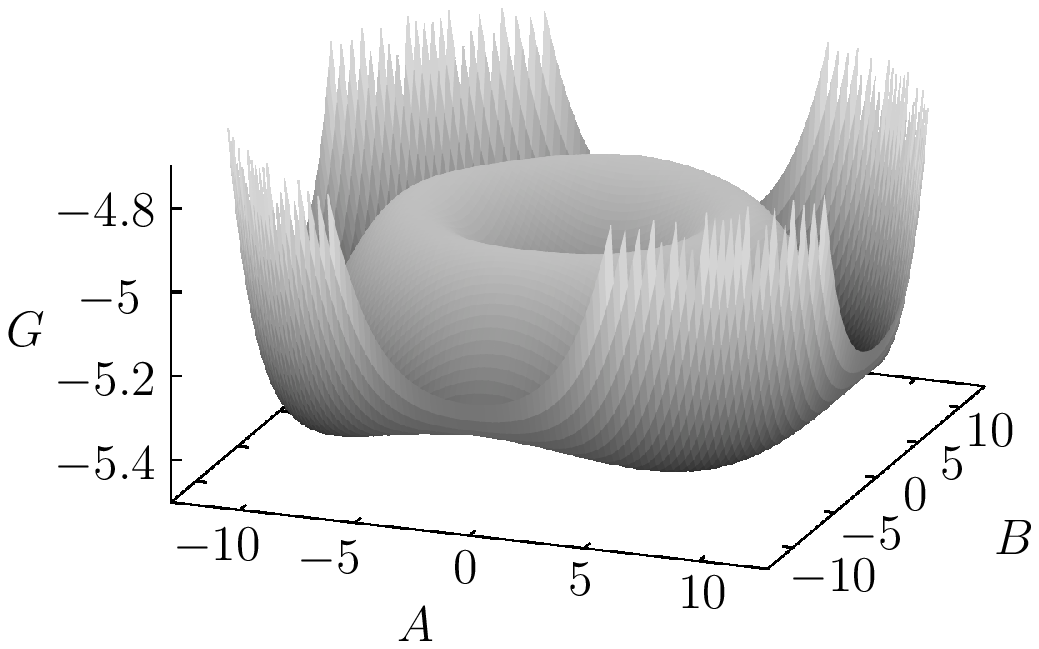}

\caption{\label{fig:G-A-B} Contour plot (left) and surface plot (right) of the reduced potential surface $G(A,B)$ (see Eq.
(\ref{eq:critical_points})) for parameters: $\epsilon=1.6\times10^{-2}$, $H=4$,
$a_{4}=6\times10^{-4}$, $a_{6}=3.6\times10^{-6}$. For these parameter, $G$ has four global minima with
$\left|A\right|=\left|B\right|$ and one local minima at $A=B=0$.  This structure with four non-trivial attractors
is due to symmetry breaking imposed by the topography $h(y)=H\cos\left(2y\right)$. }
\end{center}
\end{figure}

We observe that the term proportional to $\left(A^{2}-B^{2}\right)^{2}$ breaks the
symmetry between $A$ and $B$. Its minimization imposes that $A^{2}=B^{2}$.
Then either $A=B$, or $A=-B$. If we take into account that minimizing
with respect to $A^{2}+B^{2}$ will give only the absolute value
of $A$, we can surmise that we will have four equivalent non-trivial solutions: 
\begin{equation*}
q_{i}=-\frac{H}{3}\cos\left(2y\right)+\sqrt{2}\left|A\right|(\epsilon,a_{4},a_{6})\cos(y+\phi_{i}),
\end{equation*}
with $\phi_{i}$ taking one of the four value $\left\{ -\frac{3\pi}{4},-\frac{\pi}{4},\frac{\pi}{4},\frac{3\pi}{4}\right\} $,
with $\left|A\right|$ minimizing 
\begin{equation}
\tilde{G}(\left|A\right|)=-\frac{H^{2}}{3}+2\left(\epsilon-\frac{{a_{4}H^{2}}}{6}+\frac{5a_{6}H^{4}}{216}\right)\left|A\right|^{2}+4\left(\frac{3a_{4}}{8}+\frac{25a_{6}H^{2}}{144}\right)\left|A\right|^{4}+\frac{5a_{6}}{3}\left|A\right|^{6}.\label{eq:G-tilde}
\end{equation}

The reduced potential $G$ is plotted in figure \ref{fig:G-A-B} for
the case $\epsilon>0$ and $a_{4}>0$. The structure has four non-trivial attractors due to a breaking of the symmetry imposed by the topography $h(y)=H\cos\left(2y\right)$. For $\epsilon<0$, the minima of $G$ have the
symmetries of $h$ (potential vorticity profile have a reflexion symmetry
with respect to both $y=0$ or $y=\pi$ and an anti-reflection symmetry
with respect to both $y=\pi/2$ and $y=3\pi/2$). For $\epsilon>0$
this symmetry is broken leading to four different attractors. In figure \ref{fig:Minima-Saddle-StreamFunction}, we show the potential vorticity of two of the attractors, the corresponding saddle and the topography.

\begin{figure}
\begin{center}
\includegraphics[width=0.6\columnwidth]{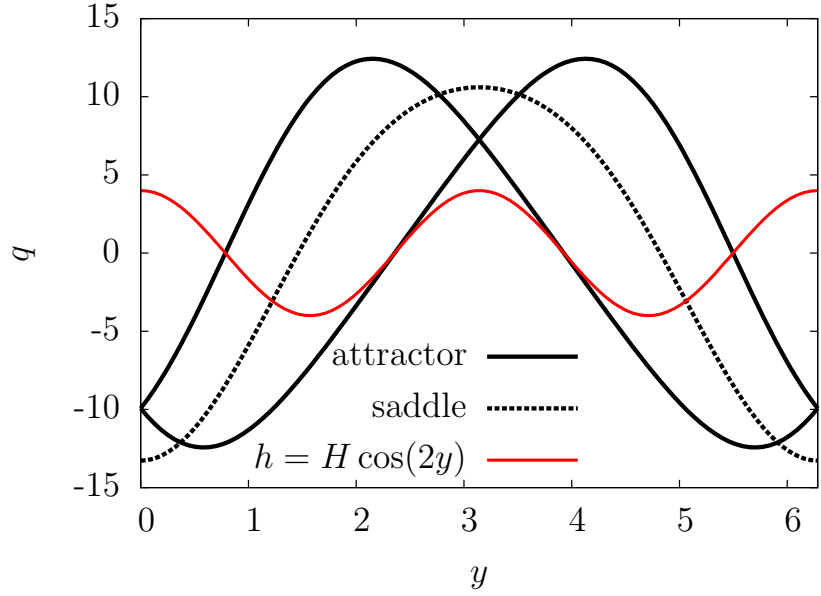} 
\caption{\label{fig:Minima-Saddle-StreamFunction}The plot depicts the topography
($h(y)=H\cos\left(2y\right)$, symmetric red curve) and two non-trivial attractors of the
potential vorticity $q$ (black solid lines) corresponding to two minima of the effective potential $G$ (see Eq. (\ref{eq:critical_points}),
and figure \ref{fig:G-A-B}) for parameter values $\epsilon>0$ and
$a_{4}>0$. Additionally, we show the saddle between the two attractors
of the effect potential $G$ (dashed black curve). }
\end{center}
\end{figure}

\begin{figure}
\begin{center}
\includegraphics[width=0.6\columnwidth]{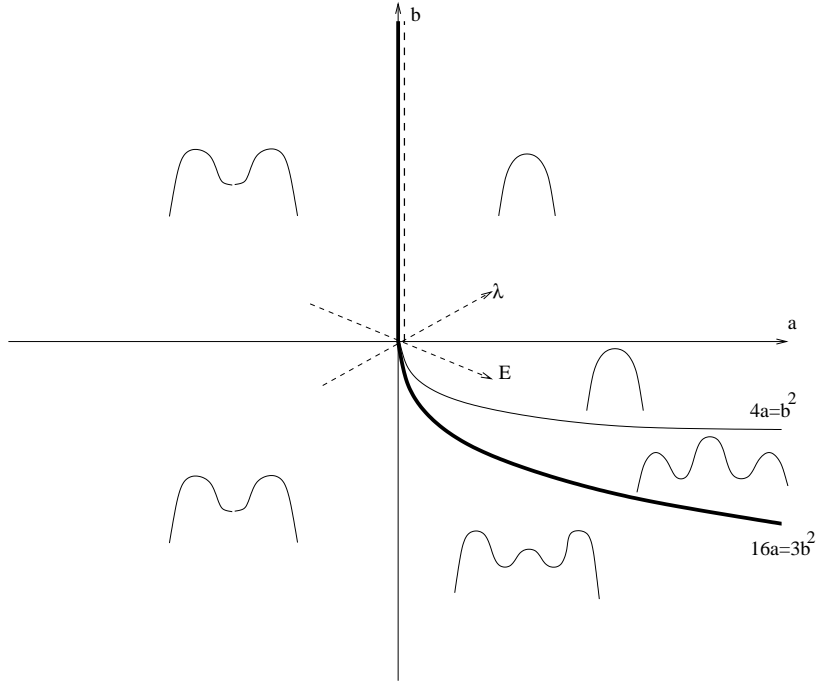}
\caption{\label{fig:Tricritique}We show the phase diagram for
a tricritical point corresponding to the maximization of the normal
form $s(m)=-m^{6}-\frac{3b}{2}m^{4}-3am^{2}$ (taken from \cite{Bouchet_Barre:2005_JSP}).
The inset show the qualitative shape of the potential $s$ when the
parameters $a$ and $b$ are changed. The black solid line corresponds to a line of first
order (discontinuous) phase transition. The black dashed line is a second order phase transition line. At the tricritical point ($a=b=0$), the first order phase transition change to a second order phase transition.}
\end{center}
\end{figure}

Considering the reduced potential $\tilde{G}$ (Eq. (\ref{eq:G-tilde})),
we recognize that the structure contains a tricritical point: a point at which
a first order transition line switches to a second order transition
line. Figure \ref{fig:Tricritique} shows a normal form for a tricritical
point. The reduced potential $\tilde{G}$ (Eq. (\ref{eq:G-tilde}))
has the same normal form structure with $a=\frac{2}{5a_{6}}\left(\epsilon-\frac{a_{4}H^{2}}{6}+\frac{5a_{6}H^{4}}{216}\right)$
and $b=\frac{8}{5a_{6}}\left(\frac{3a_{4}}{8}+\frac{25a_{6}H^{2}}{144}\right)$. 

From this last equation, we can conclude that for $a_{4}<25a_{6}H^{2}/54$
($a_{4}<0$ at leading order), we have a continuous phase transition
for $\epsilon={35a_{6}H^{4}}/{648}$ (zero at leading order).
For $a_{4}={25a_{6}H^{2}}/{54}$ ($a_{4}=0$ at leading order),
we have a tricritical point. Therefore, the transition is between a state given, at leading order, by
\begin{equation*}
q=-\frac{H}{3}\cos\left(2y\right)
\end{equation*}
to one of the four states given by
\begin{equation}
q_{i}=-\frac{H}{3}\cos\left(2y\right)+\sqrt{2}\left|A\right|(\epsilon,a_{4},a_{6})\cos\left(y+\phi_{i}\right),\label{eq:q-equilibre}
\end{equation}
where $\phi_{i}\in\left\{ -\frac{3\pi}{4},-\frac{\pi}{4},\frac{\pi}{4},\frac{3\pi}{4}\right\} $,
and $\left|A\right|(\epsilon,a_{4},a_{6})$ being the non-zero minimizer
of (\ref{eq:G-tilde}). For $a_{4}>0$ and $\epsilon$ close to zero,
we have the coexistence of both ot these states, and thus the transition
when $\epsilon$ is increased is of first order. For $a_{4}<0$
and $\epsilon$ close to zero, the transition when $\epsilon$ is
increased is a second order (continuous) transition.

\subsection{Instantons for the topography phase transition}

To summarize, we know how to describe and compute the instantons corresponding to the phase
transitions between zonal flows.
In section \ref{sec:Langevin-general} we have derived the general
theory for Langevin dynamics for field problems with potential $\mathcal{G}$, and have concluded in section \ref{sub:Instantons-Relaxation-Paths}
that instantons are the time reversed trajectories of relaxation paths for the
reversed dynamics. The corresponding equation of motion for the relaxation paths for the reversed
dynamics for the quasi-geostrophic dynamics has then been derived in section
\ref{sub:Instanton-equation-QG}. 

The general theory and Eq. (\ref{eq:relaxation-paths-QG}) show
that for the quasi-geostrophic dynamics, the reversed dynamics is
simply the quasi-geostrophic dynamics where $h$ has been replaced by $-h$
and $\mathcal{G}$ by $\mathcal{G}_{r}$, with $\mathcal{G}_{r}\left[q\right]=\mathcal{G}\left[-q\right]$.
In the example we discussed now, $\mathcal{G}$ is even (see Eq.
(\ref{eq:C-a4})) such that $\mathcal{G}_{r}=\mathcal{G}$. We remark,
 that over the set of zonal flows $\mathbf{v}=U(y)\mathbf{e}_{x}$,
the nonlinear term of the quasi-geostrophic equation
vanishes: $\mathbf{v}\left[q+h\right]\cdot\nabla q=0$.  As a consequence,
when the instanton remains a zonal flow, the fact that $h$ has to
be replaced by $-h$ has no consequence. Let us now argue that the
instanton is actually generically a zonal flow.

We assume for simplicity that the stochastic forces are homogeneous
(invariant by translation in both directions). Then $C\left(\mathbf{r},\mathbf{r'}\right)=C\left(\mathbf{r}-\mathbf{r'}\right)=C_{z}(y-y')+C_{m}(y-y',x-x')$
where 
\begin{equation*}
C_{z}(y)=\frac{1}{2\pi l_{x}}\int_{_{0}}^{2\pi l_{x}}\, C(x,y)\,{\rm d}x
\end{equation*}
 is the zonal part of the correlation function, and $C_{m}=C-C_{z}$
the non-zonal or meridional part. 

As the nonlinear term of the two-dimensional Euler equations identically vanishes,
the relaxation dynamics has a solution among the set of zonal flows.
If $C_{z}$ is non-degenerate (positive definite as a correlation
function), then relaxation paths will exist through the gradient dynamics
\begin{equation}
\frac{\partial q}{\partial t}=-2\pi\alpha l_{x}\int_{_{0}}^{2\pi}\, C_{z}(y-y')\frac{\delta\mathcal{G}}{\delta q(y')}\, {\rm d}y',\label{eq:Relaxation-Path-Zonal}
\end{equation}
where $q=q(y)$ is the zonal potential vorticity field.

Moreover, as argued in section \ref{sub:Zonal-phase-transitions},
the fact that $G$ (\ref{eq:critical_points}) is a normal form for small enough
$a_{4}$, $a_{6}$, and $H$, implies that the gradient of $\mathcal{G}$
in directions transverse to $q=A\cos\left(y\right)+B\cos\left(y\right)$ are much steeper
than the gradient of $G$. As a consequence, at leading order the
relaxation paths will be given by the relaxation paths for the effective
two-degrees of freedom $G$. Then, from (\ref{eq:Relation-G-GAB}),
(\ref{eq:critical_points}), and (\ref{eq:Relaxation-Path-Zonal})
we obtain that, at leading order, for the relaxation path given by (\ref{eq:psi-A}-\ref{eq:psi-B}), the dynamics of $A$ and $B$ are given by 
\begin{equation*}
\frac{{\rm d}A}{{\rm d}t}=-c\frac{\partial G}{\partial A}\quad {\rm and}\quad \frac{{\rm d}B}{{\rm d}t}=-c\frac{\partial G}{\partial B},
\end{equation*}
with $c=-\alpha l_{x}\int_{0}^{2\pi}\, C_{z}(y)\cos\left(y\right)\, {\rm d} y$, where we recall that $G$ is given by Eq. (\ref{eq:critical_points}).

From this result the relaxation paths are easily computed. Using the fact that
fluctuation paths are time reversed trajectories of relaxation paths, instanton
are also easily obtained. One of the resulting relaxation paths (blue curve) and
one of the instantons (red curve) are depicted in figure \ref{fig:G-A-B-Instanton}
overlapped on the contours of the potential $G$ in the $(A,B)$-plane. The corresponding two attractor involved, together
with the saddle point and examples of two intermediate states are shown in figure \ref{fig:Instanton-IntermediateStates}.

\begin{figure}
\begin{center}
\includegraphics[width=0.6\columnwidth]{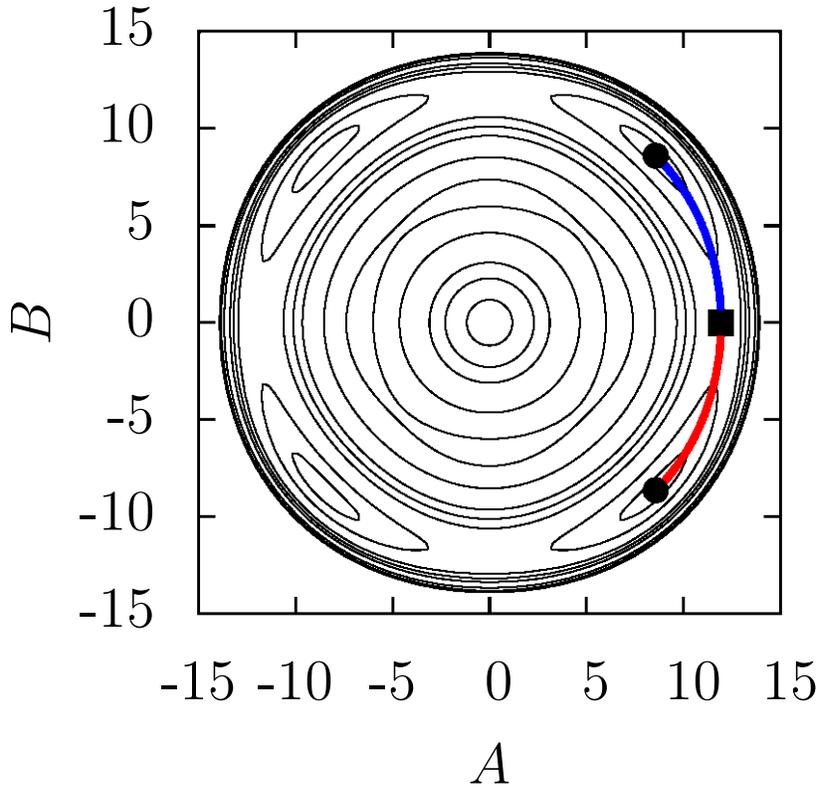}

\caption{\label{fig:G-A-B-Instanton} Contour plot of the reduced potential
surface $G(A,B)$ (same as figure \ref{fig:G-A-B}) with the superimposed
transition path between two attractors denoted by {\Large \textbullet} via a saddle $\blacksquare$.  The instanton (most probable fluctuation path from one attractor
to a saddle) is show by the solid red line, while the corresponding relaxation path from the saddle to the second attractor is given by the solid blue line.
In this case, the instanton and the relaxation paths are actually the reverse of one another.}
\end{center}
\end{figure}

\begin{figure}
\begin{center}
\includegraphics[width=0.6\columnwidth]{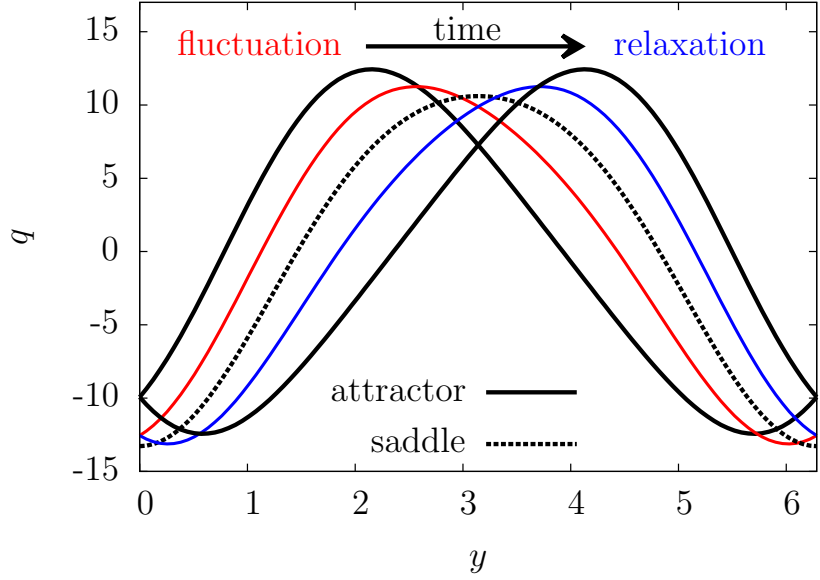}
\caption{\label{fig:Instanton-IntermediateStates} The potential vorticity
$q(y)$ for two of the non-trivial attractors (solid black curves), the corresponding saddle between the attractors (dashed black curve), and two intermediate
profiles along the instanton path (solid red curve) and the relaxation
path (solid blue curve). }
\end{center}
\end{figure}

\subsection{Dimensional analysis \label{sec:Stochastic-averaging-Reduced-Dynamics}}

In this section, we briefly discuss dimensional analysis for the
dynamics (\ref{eq:beta-plane}), with topography $h=H\cos\left(2y\right)$, and
potential $\mathcal{G}$ given by Eqs. (\ref{eq:Potential-Zonal-Transitions-1}-\ref{eq:C-a4}).
We recall these equations for clarity:
\begin{eqnarray}
\frac{\partial q}{\partial t}+\mathbf{v}\left[q-H\cos\left(2y\right)\right]\cdot\nabla q & = & -\alpha\int_{\mathcal{D}}\, C({\bf r}-{\bf r}')\frac{\delta\mathcal{G}}{\delta q({\bf r}')}\,{\rm d}{\bf r}'+\sqrt{2\alpha\gamma}\eta,\label{eq:beta-plane-1}\\
\mathbf{v}=\mathbf{e}_{z}\times\nabla \psi,\quad \omega&=&\Delta\psi, \quad q=\omega+H\cos\left(2y\right),
\end{eqnarray}
with 
\begin{equation}
\mathcal{G}=\int_{\mathcal{D}}\frac{q^{2}}{2}-a_{4}\frac{q^{4}}{4}+a_{6}\frac{q^{6}}{4}\, {\rm d}{\bf r}-\left(1-\epsilon\right)\mathcal{E}.\label{eq:Potential-Zonal-Transitions-1-1}
\end{equation}

First, let us discuss a set of convenient non-dimensional units for
our problem. We express length in units of the domain size.
The dynamics involve the following parameters $\alpha$ ($s^{-1}$), $\gamma$
($s^{2}$), $H$ ($s^{-1}$), $a_{4}$ ($s^{2}$), $a_{6}$ ($s^{4}$),
$\beta$ or $\epsilon$ (no dimension), the aspect ratio $l_{x}$
(no dimension), and the force spectrum $C$ (no dimension), energy $\mathcal{E}$
($s^{-2}$), and Casimirs $\mathcal{C}$ ($s^{-2}$). We are interested mainly in
the range of parameter for which the dynamics is bistable. Moreover,
it will be especially useful to consider the perturbative regime close
to the bifurcation described in section \ref{sub:Zonal-phase-transitions}.
As a consequence, we choose $\epsilon\ll1$, $a_{4}>0$ and $a_{4}$
sufficiently small (as discussed below), and $H$ sufficiently small
($a_{4}H^{2}\ll\epsilon$ and $a_{6}H^{2}\ll a_{4}$) such that the
phase transition is close to the one occurring for $H=0$. We recall
that these assumptions are made in order to get an explicit description
of the phase transitions, however it is important to understand that
the theory that predicts the transition rates and the instantons
does not depend on these assumptions and that the same phenomenology
will remain valid beyond this perturbative regime.

As discussed in section \ref{sub:Zonal-phase-transitions}, with these
hypotheses, the lower values of $\mathcal{G}$ are approximated by
the normal form $G$ (\ref{eq:critical_points}). From (\ref{eq:critical_points}),
we conclude that if we assume $\epsilon a_{6}\ll a_{4}^{2}$, then
the order of magnitude of $A$, the amplitude of the large scale
mode, is $\left(\epsilon/a_{4}\right)^{1/2}$. As we have chosen $a_{4}H^{2}\ll\epsilon$,
the correction due to the topography is of sub-leading order (see Eq.
(\ref{eq:q-equilibre})). The kinetic energy of the largest scale mode
is then of the order $\epsilon/a_{4}$. Subsequently, we choose $\left(a_{4}/\epsilon\right)^{1/2}$
as a time unit. We denote $H'=\left(a_{4}/\epsilon\right)^{1/2}H$, $\gamma'=\left(a_{4}/\epsilon\right)^{3/2}\gamma$,
$\alpha'=\left(a_{4}/\epsilon\right)^{1/2}\alpha$, $a'_{6}=(\epsilon/a_{4})^{2}a_{6}$,
and $q'=\left({a_{4}/\epsilon}\right)^{1/2}q$ to be the dimensionless variables in this time unit. Therefore,
we can write the non-dimensional equations, dropping the prime variables as
\begin{eqnarray}
\frac{\partial q}{\partial t}+\mathbf{v}\left[q-H\cos\left(2y\right)\right]\cdot\nabla q & = & -\alpha\int_{\mathcal{D}}\, C({\bf r}-{\bf r}')\frac{\delta\mathcal{G}}{\delta q({\bf r}')}\,{\rm d}{\bf r}'+\sqrt{2\alpha\gamma}\eta,\label{eq:beta-plane-1-1}\\
\mathbf{v}=\mathbf{e}_{z}\times\nabla\psi,\quad \omega&=&\Delta\psi, \quad q=\omega+H\cos\left(2y\right),
\end{eqnarray}
with 
\begin{equation}
\mathcal{G}=\int_{\mathcal{D}}\,\frac{q^{2}}{2}-\epsilon\frac{q^{4}}{4}+a_{6}\frac{q^{6}}{6}\, {\rm d}{\bf r}-\left(1-\epsilon\right)\mathcal{E}.\label{eq:Potential-Zonal-Transitions-1-1-1}
\end{equation}

Within these non-dimensional variables, $\epsilon$ controls the distance
to the bifurcation. The approximation of the large scale dynamics
by a few number of modes will then be valid for $\epsilon\ll1$, and the approximation
that the topography is a second order effect is controlled by $H^{2}\ll\epsilon$
and $a_{6}H^{2}\ll1$ (this also implies $a_{6}H^{4}\ll\epsilon$).

We now give a qualitative picture of  the dynamics. Recall that the stationary distribution of the stochastic process
is given by $P_{s}=Z^{-1}\exp\left(-\mathcal{G}/\gamma\right)$. The gradient
of $\mathcal{G}$ in the directions which are transverse with respect
to the modes $A\cos\left(y\right)+B\sin\left(y\right)$ is of order one, whereas the stochastic
force is multiplied by $\gamma^{1/2}$. As a consequence, typical
values of fluctuations for the stationary measure in these transverse
directions are of order $\gamma^{1/2}$. Finally, the non-dimensional
parameter $\alpha$ controls the relative order of magnitude of the
inertial (or Hamiltonian) part of the dynamics, compared to the dissipative
gradient terms in (\ref{eq:beta-plane-1-1}).

\section{Conclusions and perspectives\label{sec:Conclusions}}

We have defined Langevin dynamics for two-dimensional and quasi-geostrophic
turbulent flows. These dynamics have an energy-Casimir invariant measure.
The dissipative part of the dynamics derives from a potential that
is transverse to the Hamiltonian part of the dynamics. Moreover, the
noise autocorrelation function is the same as the kernel defining
the dissipative part. Under these hypotheses, the action is modified in a simple manner under time reversal. It is either symmetric leading to
detailed balance, or leads to a dual action which describes dynamics
that belong to the same family of physical model. These symmetries
put these Langevin dynamics in the framework of classical Langevin
dynamics. For instance, fluctuation paths are time reversed trajectories of relaxation
paths of the dual dynamics. This gives a very simple characterization
of fluctuation paths, of large deviations, and of large deviation
paths, when they exist.

We have proposed and analyzed cases with phase transitions, both continuous
and discontinuous, and of a tricritical point. This opens the study
to a rich phenomenology of processes, including bistable situations.
These Langevin dynamics with exact theoretical prediction will be very useful benchmarks for future tests of numerical
algorithms aimed at computing large deviations in turbulence problems. 

Several interesting concepts could be developed in the
future. These Langevin dynamics, give examples of turbulence problems
for which the recent results of stochastic thermodynamics could be
extended, e.g. it would be very interesting to study Gallavotti-Cohen
fluctuation relations \cite{Gallavotti_Cohen_1995_PhRvL}, or entropy
production \cite{Baiesi_Maes_2005enstrophy,jaksic2013large} in this setup. The
temporal response of the system to external driving or change of parameters
could also be studied in relation to recently studied non-equilibrium
linear response for Markovian dynamics \cite{Baiesi_Boksenbojm_Maes_Wynants_2010nonequilibrium,Mae2}.\\

Let us come back to two important and related issues not discussed
in this paper. Firstly, is it possible to give a clear mathematical
meaning to the Langevin dynamics (\ref{eq:beta-plane}), given that it may involve very rough forces
through the noise term? Or of smooth noise combined with very
weak friction? Secondly, for the dynamics (\ref{eq:beta-plane}), will large deviation results (\ref{eq:Large_Deviations})
be valid? In order to discuss these two questions, let us consider
a special case of Langevin dynamics (\ref{eq:beta-plane}), with
$C({\bf r},{\bf r}')=\Delta\delta(\mathbf{r}-\mathbf{r}')$, corresponding
to the enstrophy ensemble (see section \ref{sub:Energy-enstrophy-physicaldiscussion}).
From a physical point of view, it has been identified
for a long time that the dynamics can not be given a simple physical
interpretation. Indeed, for the enstrophy measure, the expectations
of both the energy and enstrophy are infinite. Even the expectation
for the velocity field is not defined, and most of the realization
do not lead to a physical velocity field. This is related to some
of the mathematical results in \cite{bessaih2012invariant}. These
remarks give a negative answer to the first question. Still, it has
been observed \cite{Bouchet_Laurie_2012_Onsager_Japan} that, at a
formal level, the minimization of the action can be computed explicitly
and leads to a quasi-potential which is indeed the enstrophy as may
have been expected. A natural physical question is then to understand
what happens if the noise is regularized at a scale $\delta$, much
smaller than the domain size. Recently, we have been aware of the work by Brzezniak, Cerrai and
Freidlin \cite{Brzezniak_Cerrai_Freidlin_2014quasipotential}, that actually considers
this problem. Their mathematical result, is that for any finite $\delta$
the dynamics are well defined. Moreover, that for any finite $\delta$,
a large deviation principle for exit times from a bounded domain holds
when the noise amplitude goes to zero (when $\gamma$ goes to zero
in our notation,  see Eq. (\ref{eq:beta-plane})). These large
deviations are actually described by the minimization of the action functional
(\ref{eq:Lagrangian}-\ref{eq:A}), with a kernel $C_{\delta}$ taking
into account the noise regularization. When $\delta$ goes to zero,
the large deviation functional and the minimizers of the actions actually
converge to the one corresponding to the enstrophy ensemble \cite{Brzezniak_Cerrai_Freidlin_2014quasipotential}.
These results justify the formal computation in \cite{Bouchet_Laurie_2012_Onsager_Japan},
and equivalent results would justify the formal computations presented in this
current work. However, we stress that for these results to hold, the order
of the limits ($\gamma\to 0$  and $\delta\to 0$ afterwards) is crucial. 

As discussed above, for the enstrophy ensemble, it is necessary to
regularize the noise first in order to obtain meaningful dynamics.
However, it is not yet clear which are the relavent cases, depending on the kernel
$C$ or the potential $\mathcal{G}$, when such a regularization is
necessary or not? For instance, when $a_{4}<0$ or $a_{6}>0$, see
Eq. (\ref{eq:C-a4}), such a regularization may be unnecessary, or with
a potential controlling the extremal values of the vorticity field,
such a regularization would be unnecessary. This question could be
the subject of further studies. The dynamics could also be regularized
at the level of the dissipation, for instance by adding small scale dissipation in the form of hyperviscosity
with small coefficient.

In order to conclude, we stress once more that, for applications it
would be desirable to go beyond the Langevin dynamics considered in
this paper. A first step could be for the derivation of the
slow dynamics of zonal jets in quasi-geostrophic models \cite{Bouchet_Nardini_Tangarife_2013_Kinetic_JStatPhys},
followed by large deviation computations. We consider progresses
in this direction and in other directions in future works.


\bibliographystyle{spmpsci}      
\bibliography{biblio}
%
%

\end{document}